\newif\ifsubmission\submissionfalse
\ifsubmission
\documentclass[runningheads]{llncs}
\else
\documentclass{article}
\usepackage[a4paper,margin=1.1in]{geometry}
\usepackage{mathpazo}
\usepackage[
    backend=biber,
    style=alphabetic,
    doi=false,
    maxbibnames=99
  ]{biblatex}
\fi
\usepackage{url}
\usepackage{tikz}
\usepackage{hyperref}
\usepackage{breakurl}
\usepackage{xcolor}
\usepackage{diagbox}
\definecolor{CODEorange}{rgb}{0.93, 0.4314, 0}
\definecolor{CODEanthrazit}{rgb}{0.4157,0.408, 0.435}
\hypersetup{colorlinks=true, 
                    linkcolor=CODEanthrazit,
                    citecolor=CODEorange,
                    urlcolor=CODEanthrazit}

\usepackage[T1]{fontenc}
\usepackage[utf8]{inputenc}
\usepackage{amsmath}
\usepackage{amssymb}
\newcommand{\pk}{\mathsf{pk}}
\newcommand{\sk}{\mathsf{sk}}
\newcommand{\Verify}{\mathsf{Verify}}
\newcommand{\Sign}{\mathsf{Sign}}

\newcommand{\footremember}[2]{%
    \footnote{#2}
    \newcounter{#1}
    \setcounter{#1}{\value{footnote}}%
}

\usepackage{graphicx}
\begin{document}
\usetikzlibrary{positioning, shapes, arrows.meta}
\ifsubmission
\title{Privacy-Preserving Authentication: Theory vs. Practice\thanks{This paper is based on a keynote with the same title given at the 19th IFIP Summer School on Privacy and Identity Management held between 10th and 13th September 2024 in Madrid, Spain.}}
\else
\title{Privacy-Preserving Authentication: Theory vs. Practice}
\fi
%
%

\ifsubmission
\author{Daniel Slamanig\orcidID{0000-0002-4181-2561}}
%
%
\institute{Research Institute CODE, Universität der Bundeswehr München, Germany
\email{daniel.slamanig@unibw.de}}
\else
\author{%
            Daniel Slamanig\footremember{delta}{\href{mailto:daniel.slamanig@unibw.de}{daniel.slamanig@unibw.de}. This paper is based on a keynote with the same title given at the 19th IFIP Summer School on Privacy and Identity Management held between 10th and 13th September 2024 in Madrid, Spain and appears in the Proceedings of the Summer School.}}
        \date{\normalsize Research Institute CODE, Universit{\"a}t der Bundeswehr M{\"u}nchen}
\fi
\maketitle              
\begin{abstract}
With the increasing use of online services, the protection of the privacy of users becomes more and more important. This is particularly critical as authentication and authorization, as realized on the Internet nowadays, typically rely on centralized identity management solutions. Although those are very convenient from a user's perspective, they are quite intrusive from a privacy perspective and are currently far from implementing the concept of data minimization. Fortunately, cryptography offers exciting primitives such as zero-knowledge proofs and advanced signature schemes to realize various forms of so-called anonymous credentials. Such primitives enable online authentication and authorization with a high level of built-in privacy protection (what we call privacy-preserving authentication). Though these primitives have already been researched for various decades and are well understood in the research community, unfortunately, they lack widespread adoption. In this paper, we look at the problems, what cryptography can do, some deployment examples, and barriers to widespread adoption, which we discuss using the example of the EU Digital Identity Wallet (EUDIW) and the recent discussion and feedback from cryptography experts around this topic. We also briefly comment on the transition to post-quantum cryptography.

\end{abstract}
\section{Introduction}\label{sec:intro}

Authentication and authorization are two closely coupled and important tasks that are frequently required when users want to access resources in the digital realm, e.g., services on the Internet. Loosely speaking, authentication is the task of confirming whether a user is really who they pretend to be, e.g., by demonstrating the knowledge of some secret such as a password or a secret signing key associated with the user. Today, one increasingly uses a combination of more than one factor (multi-factor authentication), like the knowledge of a password and the possession of Subscriber Identity Module (SIM) card. Authorization is the task of giving users permissions to access certain resources, i.e., deciding whether a certain authenticated user has sufficient permissions to perform some action. Henceforth, we will just talk about authentication and note that typically authorization can be performed based on information associated with the identity after authentication and without any further interaction with the user. A typical example is the use of attributes associated with users, e.g., any user with attribute \texttt{role=Student} might be granted access to a certain service. 

A convenient way to think about a digital identity is that of a set of attributes that describe a person. Now, in a typical scenario a person's identity comprises several not necessarily disjoint partial identities (cf. \cite{anonterminology}), where each of them represents a person in a specific context or role and can be thought of as a certain subset of attributes. For instance, every person has a set of attributes related to health (e.g., health status, medical history), which they might not want to include in their partial identity for ``work''. The complete identity of a person is represented by the union of all the attributes of all the partial identities of this person. Since some of these partial identities include sensitive information, e.g., the health data mentioned above, from a privacy perspective it is desirable to have full control over which attributes are revealed in a certain context. This is particularly relevant in the digital realm where it is easy to collect such information, connect it to other information sources and use it to build up ever increasing profiles of individuals.

Limiting the amount of information that is provided as much as possible is known as data minimization. It is a concept of growing importance in a globally connected digitized world and a fundamental principle behind many data protection regulations such as the General Data Protection Regulation (GDPR) in the European Union or the Health Insurance Portability and Accountability Act (HIPAA) in the United States. This helps to preserve privacy and can help to ensure as much unlinkability as
possible, making it harder to easily connect all the partial identities of a person. One issue that is important to stress is that for many services on the Internet, concretely identifying information such as \texttt{name} or \texttt{DateOfBirth} of a user might not be required in the process of authentication, e.g., it can be sufficient for an individual to demonstrate that they are old enough to consume a certain service. What, however, can be required in various cases such as governmental applications, is that actions of the same partial identity of a person can be linked over different sessions. This can be achieved via so-called pseudonyms, which represent identifiers other than the person's real name (or other uniquely identifying information) that are independent of the person and provide a sufficient degree of anonymity.\footnote{We stress that pseudonymity and anonymity are not used in the strict legal sense but rather in a technical sense relevant to the paper’s context.}

After having introduced some basic concepts, our goal is to look into how authentication on the Internet typically looks today and how it could look when putting privacy as the main design goal. We will discuss issues with these existing mechanism and present a concept from the cryptographic literature called anonymous credentials. This is an important tool to realize privacy-preserving authentication first envisioned by Chaum \cite{DBLP:journals/cacm/Chaum85} and later realized by Camenisch and Lysyanskaya \cite{DBLP:conf/eurocrypt/CamenischL01} as well as Brands \cite{BrandsBook} in the early 2000s. After around 25 years, there is a large body of research on anonymous credentials and related primitives (cf. \cite{DBLP:conf/secsr/KakviMPQ23,cryptoeprint:2023/1039}). Unfortunately, although this concept is well known in the research community, and over the recent years increasingly attracting practitioners, it still lacks a widespread deployment. Besides providing a high-level discussion of the theory of anonymous credentials, this paper will also look at the practice of anonymous authentication and potential issues that are hindering a widespread adoption. As a supporting use case, we take the European Digital Identity Wallet (EUDIW), its architectural reference framework (ARF) and the recent feedback given by cryptographic experts \cite{cryptoeudi}, which highlights obstacles on the way to deploying such a ``new'' technology into an existing (legacy) infrastructure.
~\\[0.5em]\textbf{Outline of this paper.} In Section \ref{sec:tradauth} we review traditional authentication on the Internet. Then, in Section \ref{sec:tradpriv} we introduce relevant privacy properties and discuss privacy of traditional authentication approaches. In Section \ref{sec:anontheory} we introduce the concept of anonymous credentials from a conceptual (theoretical) perspective. Then, in Section \ref{sec:anonpractice} we discuss anonymous credentials from a practical perspective and in particular implementations, standardization efforts and real-world use cases. In Section \ref{sec:euidw} we take a look at the European Digital Identity Wallet (EUDIW), its planned realization and related privacy issues. Then in Section \ref{sec:barriers} we look at barriers and issues to be considered when aiming at deploying anonymous credentials such as integrating them into the EUDIW. Finally, in Section \ref{sec:conclusion} we conclude this paper and give an outlook on future aspects related to deploying anonymous credentials.

\section{Traditional Authentication and Identity-Management}\label{sec:tradauth}

The very traditional way of authenticating users on the Internet is that every service realizes its own authentication mechanism, meaning that a user establishes a partial identity with every single service (typically pseudonyms or more likely email addresses). These partial identities are though typically not distinct, as users tend to re-use their email addresses either as an explicit pseudonym or as an attribute given to the service provider. The latter is usually required for functionality reasons, i.e., account recovery. 
\subsection{Password-based Authentication} The most common traditional authentication mechanism is via user-chosen passwords associated with user accounts. Since strong passwords are hard to remember, it is well known that users tend to choose too weak passwords \cite{DBLP:conf/sp/Bonneau12} as well as reuse them among different services \cite{DBLP:conf/soups/WashRBW16}. While there are password-management tools that help to overcome these issues, research shows that the use of such tools significantly lags behind \cite{DBLP:conf/soups/ZhangPBC19} and due to various reasons does not see widespread use. Despite all these issues, due to its simplicity and easy deployment, it is still the most common authentication mechanism. 
\subsection{Signature-based Authentication}  From a security perspective, two-factor authentication that relies on strong cryptography is more desirable and has been increasingly promoted. A popular choice is the use of the FIDO2\footnote{\url{https://fidoalliance.org/fido2/}} de-facto standard from the Fast IDentity Online (FIDO) alliance. Here one either uses a dedicated hardware token or a pure software component (typically called Passkey\footnote{\url{https://fidoalliance.org/passkeys/}}) as the second factor. The latter can be easily used with any device like a smartphone and thus enhances the usability. 

The technical protocol details are out of the scope of this paper, but the relevant part is that it uses public-key cryptography and in particular digital signatures. We recall that in the context of a digital signature scheme every user can generate a pair $(\sk,\pk)$ consisting of a secret (signing) key $\sk$ and a public (verification) key $\pk$, where $\pk$ is made public. A user knowing $\sk$ can produce a signature $\sigma \gets \Sign(\sk,m)$ for any message $m$, and using $\pk$ and, given $(m,\sigma)$, anyone can check the validity of the signature using $\{0,1\}\gets \Verify(\pk,m,\sigma)$, where $1$ indicates a valid signature. A valid signature gives the guarantees that $m$ has really been signed by the holder of the secret key corresponding to $\pk$ and has not been altered in any way. Security requires that a signature scheme is existentially unforgeable under a chosen message attack (i.e., provides EUF-CMA security). This means that when only having $\pk$ and access to signatures for arbitrarily and adaptively chosen messages, it is not possible to produce a valid signature $\sigma$ for some message $m'$ that has not been explicitly signed by the signer, i.e., for which no signature was requested.

In conventional web authentication, $\pk$ is typically just stored together with the identity and thereby establishing the binding between the individual and the key. More generally, this binding between $\pk$ and an identity is achieved via explicit certification. The most common way is to rely on a public-key infrastructure (PKI). This means that there is some trusted certification authority (CA) that signs $\pk$ together with a set of attributes of the user, resulting in a certificate that can be verified by anyone trusting the public key of the CA. This certificate is typically used for signature-based authentication when using an official governmental electronic identity (eID), as largely deployed within the European Union, where some governmental body acts as a CA. 
\subsection{Single Sign-On} 
\begin{figure}[!ht]
\centering
\begin{tikzpicture}[node distance=5cm, auto]
    \tikzstyle{user} = [circle, draw, fill=blue!30, minimum size=1.5cm, align=center] 
    \tikzstyle{rp} = [rectangle, draw, fill=gray!30, text width=3cm, align=center] 
    \tikzstyle{idp} = [rectangle, draw, fill=orange!30, text width=3cm, align=center] 

    \node[user] (user) {User};
    \node[rp, right of=user, node distance=5.5cm] (rp) {Relying Party \\ (RP)};
    \node[idp, above of=rp, node distance=4cm] (idp) {Identity Provider \\ (IdP)};

    \draw[->, >=stealth, thick] ([yshift=0.2cm]user.east) -- ([yshift=0.2cm]rp.west) node[midway, above, font=\small] {1. Request access};
    \draw[->, >=stealth, thick] ([yshift=-0.2cm]user.east) -- ([yshift=-0.2cm]rp.west) node[midway, below, font=\small] {5. Send Token};

    \draw[->, >=stealth, thick] (rp) -- (idp) node[midway, right=0.1cm, font=\small] {2. Redirect to IdP};

    \draw[->, >=stealth, thick] ([xshift=0.3cm]idp.south) -- ([xshift=0.3cm]user.north) node[midway, right=0.2cm, font=\small] {4. Issue Token};
    \draw[<-, >=stealth, thick] ([xshift=-0.2cm]idp.south) -- ([xshift=-0.2cm]user.north) node[midway, left=0.3cm, font=\small] {3. Authenticate};

\end{tikzpicture}
\caption{Abstract concept of single sign-on.}
\label{fig:sso}
\end{figure}
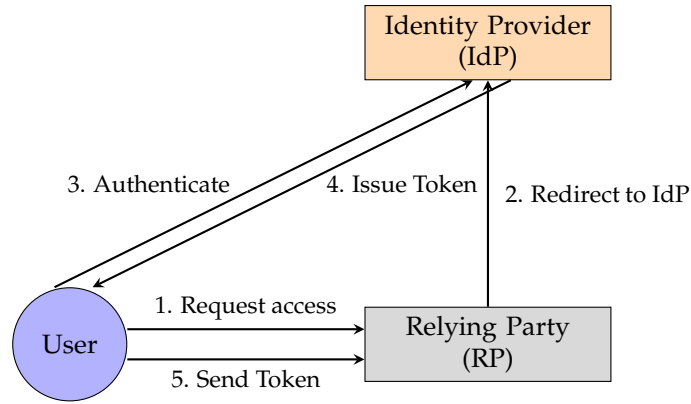

A widespread concept used nowadays is that of single sign-on (SSO), a concept with a rich history and prominent schemes such as Microsoft Passport (cf. \cite{sso-pashalidis} for a comprehensive treatment of that topic). It makes account management simpler as it requires users to authenticate only at a single centralized entity (typically a large technology company such as Google, Microsoft, or Meta) acting as an identity-provider (IdP) for the user. Then one employs an authentication and authorization layer such as OpenID Connect (OIDC), which enables authentication at third-party services (so-called relying parties or RPs) via the IdP. We illustrate the process in Figure \ref{fig:sso}.

Essentially, whenever a user wants to authenticate at an RP (1), the user is re-directed to its IdP (2) where the user actually authenticates using e.g., password- or signature-based authentication (3). After a successful authentication, the IdP generates a token (an ID Token in case of OIDC), signed by the IdP (4), which the user can hand over to the RP (5) as a proof that the user has authenticated to the IdP.

\section{Privacy Properties and Traditional Approaches}\label{sec:tradpriv}

Now having discussed ways of authenticating users on the Internet, we want to discuss privacy properties that might be desirable in such mechanisms. 

\subsection{Privacy Properties}

We will briefly summarize the privacy properties that are most important in the context of authentication at an intuitive level. We will not take into account other important security properties of authentication mechanisms which are not relevant to privacy, e.g., that it is hard to impersonate other users. Also in our further consideration we will not consider privacy on the network layer, i.e., identifiers such as IP addresses or other metadata on the network layer that might be used to identify users. There are overlay networks  \cite{DBLP:conf/sp/SyversonGR97,DBLP:conf/uss/DingledineMS04} and in particular Tor\footnote{\url{https://www.torproject.org}} to achieve anonymity. They can be used to hide such information and obtain anonymity on the network layer in practice. For simplicity, we just assume that using such techniques, anonymity is realized in an ideal way. Yet we stress that in practice for such networks to be effective there are many issues that need to be considered \cite{DBLP:conf/diau/Raymond00}. Now we list the properties: 

\paragraph{\bf Selective Disclosure.} It should be possible to let a user decide which information about their identity (e.g., the concrete attributes) will be revealed to a RP and thus adhere to the principle of data minimization. 

\paragraph{\bf Unlinkability.} Technically, unlinkability means that different actions (or transactions) of a user cannot be linked together. We will use the terms IdP, even though it might be a certificate issuer only, and RPs as a generic term for service providers. Since in such a setting there are different parties involved, we have to consider unlinkability in settings where several of these parties collaborate in trying to break the unlinkability. Below, we will use the terminology in \cite{cryptoeudi}: 

\begin{description}

\item[Unlinkability with respect to RPs:] If a user authenticates to different RPs, those RPs cannot determine whether these transactions correspond to the same or two distinct users. Thereby, RPs may have access to additional auxiliary information that might help to correlate the transaction data. Unlinkability should also hold for the same RP, i.e., when a user authenticates multiple times to the same RP—of course, unless the user has indeed established a unique pseudonym that would intentionally break unlinkability.

\item[Unlinkability with respect to IdP:] The IdP should not learn any information about which RPs a user authenticates to. So even if the IdP observes all actions at the RPs, it should not be able to learn such information. This property is sometimes called unobservability.

\item[Unlinkability with respect to IdP and RPs:] The above two flavors of unlinkability consider each party being malicious in isolation. A stronger and actually more realistic version of unlinkability considers a setting where the IdP and RPs can be malicious and collude, i.e., the IdP and the RPs can bring all their information together. Still it should not be possible to track and re-identity users. This property is sometimes called untraceability. 
\end{description}
\textbf{What does this mean for anonymity?} We stress that unlinkability is a technical term and needs to be rigorously formalized to capture the desired intuition. For instance, it is clear that unlinkability always needs to be considered along with the revealed attributes and it can only be guaranteed among a set of users with the same attributes, i.e., the set in which the identity of the respective user is hidden. The degree of anonymity provided in such a setting always depends on the size of this so called anonymity set, i.e., the set of users that could reveal the same attributes. Thus, a larger anonymity set implies stronger anonymity and the more unique the combination of revealed attributes, the smaller the degree of anonymity. Technically unlinkability can thus still be satisfied if there are at least two potential users, but for a small anonymity set the degree of anonymity will be low.

\subsection{Discussion of Privacy in Traditional Approaches}
Subsequently, we briefly discuss privacy in context of traditional approaches that are currently deployed in practice.
~\\[0.5em]\textbf{Password-based authentication.} In the classical password-based authentication approach, every RP maintains its own system and simultaneously plays the role of the IdP (there is no external IdP). Consequently, there are no different flavors of unlinkability and different RPs can link any transactions of users when collaborating as long as there is sufficient information to link (e.g., common pseudonyms, email addresses, etc.). From a selective disclosure perspective, all the attributes that are revealed once to the RP (during registrations or actions) need to be considered as ``always revealed''. Data minimization thus can only be realized when the mandatory information required by the RP just represents the minimal information necessary. 
~\\[0.5em]\textbf{Signature-based authentication.} For a signature-based approach, just like in password-based authentication, the (required) user attributes are just stored with each RP. Besides, in this case, the certificate — which is required for authentication to check the authenticity of the public key — always reveals all attributes encoded within it.
~\\[0.5em]\textbf{Single sign-on.} For the single sign-on approach, as it is realized by OpenID Connect, we can observe the following. The IdP knows all the required attributes of the user, and when authenticating at an RP the IdP may only reveal attributes (via the ID token) that are required. But this is clearly outside the control of the user and so selective disclosure is not strictly enforced. When it comes to unlinkability, different RPs can typically track a user as the ID token contains some unique identifier of the user. The IdP clearly can link all interactions as it is the one issuing the ID token. Moreover, any collaboration between the IdP and RPs can trivially link actions of one user together.
~\\[0.5em]\textbf{Beyond the currently deployed approaches.} We want to mention that Kim Cameron with his Laws of Identity\footnote{\url{https://www.identityblog.com/stories/2005/05/13/TheLawsOfIdentity.pdf}} and the design of Windows CardSpace (Infocard) in the early 2000s already advocated and pushed for integrating of user-controlled selective disclosure into SSO solutions. Moreover, while not (yet) practically deployed, we want to note that there are recent research works, e.g., by Lehmann et al. \cite{DBLP:journals/popets/KroschewskiL23,cryptoeprint:2024/1124}, that investigate how currently used approaches like OpenID Connect could work with an increased privacy protection. Also, Yeoh et al. \cite{DBLP:conf/uss/YeohKHKH23} present an approach to increase privacy in the FIDO2 protocol. We will not further discuss these recent approaches here but note that they use techniques, in particular anonymous credentials, that we are going to discuss subsequently.

\section{Anonymous Credentials: The Theory}\label{sec:anontheory}

We now first want to clarify what we mean by ``privacy-preserving authentication''. Essentially, this is a signature-based authentication approach that supports $i)$ selective disclosure and $ii)$ unlinkability in all the flavors that we have discussed. Consequently, even if IdP and RPs collude, it is not possible to track users. Note that this makes the user decide which information (e.g., attributes) are revealed and provide a strong unlinkability guarantee by default. As already mentioned in Section~\ref{sec:intro}, there are different types of anonymous credentials with a set of different features as well as (closely) related primitives (cf. \cite{DBLP:conf/secsr/KakviMPQ23,cryptoeprint:2023/1039}). In the following we will consider a simple abstract concept and will then discuss different (basic) features in Section \ref{sec:abcfeatures}.

\subsection{The Abstract Concept}
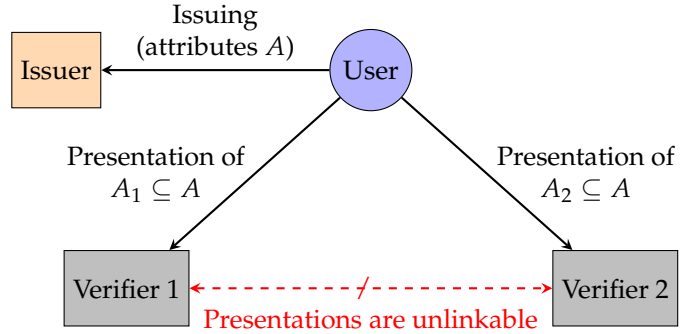
\begin{figure}[ht!]
    \centering
\begin{tikzpicture}[node distance=2cm, every node/.style={align=center}]

\tikzset{
    user/.style={circle, draw, fill=blue!30, minimum size=1cm},
    issuer/.style={rectangle, draw, fill=orange!30, minimum size=1cm},
    verifier/.style={rectangle, draw, fill=gray!50, minimum size=1cm},
    credential/.style={draw, dashed, ->, >=stealth, thick},
    showing/.style={draw, <-, >=stealth, thick},
    showing2/.style={draw, ->, >=stealth, thick},
    unlinkable/.style={draw, <->, dashed, >=stealth, thick, red},
    strikeout/.style={draw, thick, black},
}

\node[issuer] (issuer) {Issuer};
\node[user, right=3cm of issuer] (user) {User};
\node[verifier, below left=2cm and 2cm of user] (verifier1) {Verifier 1};
\node[verifier, below right=2cm and 2cm of user] (verifier2) {Verifier 2};

\draw[showing] (issuer) -- node[midway, above] {Issuing \\(attributes $A$)} (user);

\draw[showing2] (user) -- node[midway, left] {Presentation of \\$A_1\subseteq A$} (verifier1);
\draw[showing2] (user) -- node[midway, right] {Presentation of \\$A_2\subseteq A$} (verifier2);

\draw[unlinkable] (verifier1) -- ++(2.5,0) coordinate(temp1)  |-  (verifier2) node[pos=0.6,red,sloped] {$/$};

    \node[below=2.5cm of user] (unlinktext) {\textcolor{red}{Presentations are unlinkable}};

\end{tikzpicture}
    \caption{Abstract concept of an anonymous credential system.}
    \label{fig:AnonCreds}
\end{figure}

Abstractly, anonymous credentials involve three types of entities: users, issuers (IdPs) and verifiers (RPs). For simplicity let us assume that we have only a single issuer in the system. The protocol flow is as follows (see Figure \ref{fig:AnonCreds}): A user obtains a credential for an attribute set $A$ from an issuer via an interactive issuance protocol, e.g., a user might obtain a governmental ID containing name, date of birth, nationality, etc.
At the end of this interaction, the user has a credential, which can be seen as a signature of the issuer on attributes $A$. Then, the user can perform a presentation protocol with verifiers, where in each interaction a subset $A_i\subseteq A$ of the attributes in the credential can be revealed. For simplicity, here we use a simple selective disclosure of attributes, but this could actually be a more complex predicate over a subset of attributes (as discussed in Section \ref{sec:abcfeatures}). For instance, a user might want to consume some content online where they need to demonstrate that they satisfy some age rating or that they are residents of some country. Note that in those cases revealing the full identity of the user is typically not necessary. If a verifier accepts such a presentation protocol, then it can be sure it interacts with a user who has a valid credential from the issuer for the subset of attributes that were revealed during the presentation protocol. Note that as long as none of the attributes presented are identifying, and there is more than one user who could potentially also reveal the same set of attributes, the user stays anonymous. Note that in Figure \ref{fig:AnonCreds} we can also have that both verifiers are the same entity. 
~\\[0.5em]\textbf{Security.} What we request from an anonymous credential scheme is that it is \emph{unforgeable}, that is, it is not possible for a user to demonstrate some set of attributes for which they did not obtain a credential. Moreover, the scheme is \emph{unlinkable}, which means that it is not possible to link together the presentations of the same user. We note that for any candidate scheme considered to be deployed in practice it is imperative to come with a rigorous proof of security in a realistic security model.
~\\[0.5em]\textbf{Extensions to this simple model.} As briefly mentioned above, realistic scenarios often involve multiple issuers — a situation sometimes referred to as a multi-authority anonymous credential system. In such a setting, every issuer has an independent signing key to issue credentials, and a presentation might then involve more than one credential. Such an extension is typically straightforward, but if one targets at specific properties, e.g., constant-size credential presentations irrespective of the number of issuers, dedicated constructions exist \cite{DBLP:conf/scn/HebantP22,DBLP:conf/ccs/MirBGLS23}. 

Another flavor of having multiple issuers is that of threshold issuance anonymous credentials \cite{DBLP:conf/ndss/SonninoABMD19,DBLP:conf/sp/DoernerKLST23}. Here, rather than every issuer holding an independent signing key, all the credentials are issued under a single signing key. However, the key is shared among all the issuers using a $(k,n)$-threshold secret sharing technique. In such a setting, there are $n$ issuers each holding a share of the secret key and $k$ issuers need to collaborate to be able to issue a credential. Note that this is a means to distribute the trust in the issuers and adds security, as it is required that even leaking $t<k$ secret key shares does not allow one to learn the signing key. Moreover, it adds robustness as the availability of any $k$ out of the $n$ issuers is sufficient to issue credentials.

\subsection{Anonymous Credentials and Distributed Ledger Technologies}
We note that today there is a trend to move away from a centralized setting towards a decentralized identity. A popular concept in this decentralized identity space is that of self-sovereign identity (SSI), with Sovrin\footnote{\url{https://sovrin.org}} being a
prominent example that leverages distributed ledger technologies (DLTs) and concepts from this domain such as decentralized identifiers (DIDs).

When talking about the use of concepts from DLTs together with anonymous credentials, there are two different approaches that should not be confused. First, proposals such as decentralized anonymous credentials \cite{DBLP:conf/ndss/Garman0M14} or zk-creds \cite{DBLP:conf/sp/RosenbergWGM23}, that crucially use DLTs to realize novel types of anonymous credentials. This is done with the goal of not requiring explicit credential issuers \cite{DBLP:conf/ndss/Garman0M14}, or they are designed to solely rely on existing identity documents and thus do not require issuers to issue additional signature-based credentials \cite{DBLP:conf/sp/RosenbergWGM23}.

Second, the practical implementation of existing anonymous credentials using technologies provided by DLTs such as DIDs. For instance, in the context of anonymous credentials, DIDs commonly provide a registry-based mechanism to ``anchor'' the identities of credential issuers, holders, and verifiers. Anonymous credential solutions might also leverage DLT to store credential schemas and registries for credential revocation. Consequently, even though DLTs and DIDs can function independently of anonymous credentials (and vice versa), they are often tightly intertwined in practice, and many standardization initiatives have evolved in parallel across both areas. Examples are Hyperledger AnonCreds developed alongside Hyperledger Indy, and the World Wide Web Consortium (W3C) Verifiable Credentials standards developed in tandem with DID standards (cf. Section \ref{sec:anonpractice} for more details). 

Moreover, the European Union Digital Identity Wallet (EUDIW) later discussed in Section \ref{sec:euidw}, although not related to DLTs and in contrast to Sovrin not yet related to anonymous credentials, can be viewed as an approach towards SSI. At a high level, users are collecting certified attributes, i.e., verifiable credentials, from different sources and then presenting (subsets of) verifiable credentials from this collection.

\subsection{Features of Anonymous Credentials}\label{sec:abcfeatures}
In the following, we present some basic, distinguishing features of anonymous credentials, as well as a list of extended features. We try to be comprehensive in our presentation, but do not claim to be exhaustive. See also \cite{DBLP:journals/istr/CamenischDELNPP14} for a discussion of features.
~\\[1em]\textbf{Basic features.} Some basic and distinguishing features of anonymous credentials are as follows:
\begin{description}
    \item[Single-use vs. multi-use credentials.] In the single-use case, if a credential is used more than once, different presentations become linkable. Multi-use credentials support an unlimited number of unlinkable presentations. When talking about anonymous credentials, one typically implicitly assumes multi-use ones.
    \item[Support of attributes.] Typically it should be possible to encode attributes into an anonymous credential. However, over the recent years some use cases appeared in industry in which attributes are not really required. The latter concept is then typically called anonymous tokens, starting with Privacy Pass \cite{DBLP:journals/popets/DavidsonGSTV18}, which are built from blind signatures or related primitives such as oblivious pseudo-random functions (OPRFs). 
    \item[Non-transferability.] An issue with any digital credential is that it can easily be copied and thus illegally distributed. This gets particularly delicate in case of privacy-preserving authentication mechanisms where the identity will not be revealed. While there is no panacea, a typical approach is to bind it to hardware \cite{DBLP:conf/crypto/Brands93,DBLP:conf/ccs/HanzlikS21}, as done within the related primitive of direct anonymous attestation (DAA) \cite{DBLP:conf/ccs/BrickellCC04} built within every Trusted Platform Module (TPM), available in most modern computers nowadays. An alternative approach is to force users to share a valuable secret whenever they share their credentials \cite{DBLP:conf/eurocrypt/CamenischL01}, a concept which, however, seems rather hard to implement in practice. We stress that even when credentials are bound to hardware and thus a device, this does not yet prevent the sharing of the device, e.g., lending it to a family member or friend. There are recent approaches to involve biometrics to prevent this problem. Hesse et al. suggest to use a dedicated hardware token that displays a photo of the credential holder and which is bound to an anonymous credential \cite{DBLP:conf/uss/HesseSS23}. Adams suggested to directly integrate a biometric template as attribute into the anonymous credential \cite{DBLP:journals/scn/Adams11}. While this approach has recently been shown to be practical \cite{DBLP:journals/compsec/RodriguezKS24}, it is best suited for physical access control (or requires additional trust assumptions).
    \item[Expressiveness of attribute presentations.] Credentials may either only allow to reveal or withhold attributes (this can be thought of redacting the non-presented attributes) or be able to prove arbitrary statements about attributes encoded in the credential, e.g., attribute \texttt{country} represents any of the EU countries and attribute \texttt{birthdate} is so that the holder is above 18 years old. The latter requires the use of zero-knowledge proofs (cf. Section \ref{sec:constrprinc}), while from a construction perspective former can be realized without them. 
    \item[Public vs. private verification.] Typically, credentials are publicly verifiable using the public key of the issuer. However, there are also constructions \cite{DBLP:conf/ccs/ChaseMZ14,DBLP:journals/iacr/Orru24} and use cases, e.g., private groups in the Signal messenger \cite{DBLP:conf/ccs/ChasePZ20}, where the issuer is identical to the verifier and verification uses the secret of the issuer. This can allow for more efficient constructions that can avoid the use of bilinear groups and thus specific pairing-friendly elliptic curves (cf. Section \ref{sec:hwcomp} for a discussion). 
\end{description}
\noindent\textbf{Extended features.} Moreover, there are a number of additional features that can be highly relevant depending on the application.
\begin{description}
    \item[Revocation.] As in conventional public-key infrastructures (PKIs), one typically requires a means to invalidate already issued credentials before their expiration. However, when desiring unlinkability this cannot be a simple serial number that is always checked and consequently revocation in such a scenario is more complex and adds additional overhead. While this overhead is typically manageable from a computational perspective \cite{DBLP:conf/cms/LaponKDN11} it might introduce additional issues and needs to be thoroughly evaluated. Loosely speaking, it typically mirrors the concept of a credential revocation list, but in a privacy-preserving way. One trivial approach that avoids such mechanisms and is conceptually simple is frequent re-issuing of credentials to users. This, however, adds a significant overhead on the issuer and puts some online requirements on the user. Moreover, this can be dangerous from a privacy perspective, as a time-based correlation between issuing directly before using can drastically reduce the size of the anonymity set.
    \item[Robustness.] Devices storing credentials can be lost or stolen. So it is important to make credentials recoverable or robust to device loss and unavailability respectively. One strategy is to back up credentials in a way so that they can be recovered when the device is not available anymore \cite{DBLP:conf/acns/BaldimtsiCHKLN15}. An alternative approach is to consider sharing the credential among more than one device (in a threshold way) and consider threshold presentation protocols for credentials \cite{cryptoeprint:2024/1874}, i.e., $t$ out of $n$ devices need to be available. To the best of our knowledge this issue is rarely discussed, but we believe that it is an important one that needs to be considered for practical deployment.
    \item[Delegation.] A user might require to share access to resources and services with another person or among their different electronic devices. In practice, credentials are usually issued in a hierarchical manner, e.g., in a PKI there is a chain of certificates between the user certificate and a trusted root certification authority. In case this chain of issuers (or delegators) reveals sensitive information about the issuer's organizational structure or the credential holder, one can rely on delegatable anonymous credentials \cite{DBLP:conf/crypto/BelenkiyCCKLS09} (cf. \cite{DBLP:journals/popets/MirSBM23} for a recent discussion on practically efficient schemes).
    \item[Pseudonyms.] In some use cases it is required that while users are not identifiable, their action can still be linked together (even in case of multi-use credentials). Whenever this is required, it is possible to use context-specific pseudonyms. Here, from a unique value encoded in the credential (e.g., from its secret key) and a given context (string), a user always deterministically derives the same pseudorandom identifier that can be used as a pseudonym in this context. For instance, all actions of a user in the health domain are linkable, but are unlinkable to other domains.
    \item[Issuer-hiding.] In a setting where a credential could have been obtained from multiple issuers, e.g., one country within the EU, it might be privacy critical to reveal the concrete issuer. The same holds when presenting multiple credentials from several issuers, where the combination of issuers might already leak a lot of information. Consequently, the issuer-hiding concept \cite{DBLP:conf/cans/BobolzEKRS21,DBLP:conf/pkc/Connolly0P22} allows to not reveal the issuer of a credential to the verifier during presentation. Rather than revealing the issuer, the user just demonstrates that an “issuer-policy”, i.e., membership in a set of issuers, which is defined by the verifier, is satisfied.
    \item[Blind issuing of attributes.] An issuer might be required not to learn certain attributes when issuing a credential. For instance, users typically need to include a secret key as an attribute into the credential, but the issuer is only allowed to get to know the public key. Moreover, it might be the case that the user wants to transfer a hidden attribute from one credential to another newly issued one without revealing anything about it (not even a function as in case of the secret and public key) to the issuer.
    \item[Inspection.] Certain applications might require to make anonymity conditional and to escrow identifying information with the presentation of a credential (e.g., by encrypting it for a certain party). Consequently, this information can be opened by a third party when required and thus enables re-identification of the otherwise anonymous user.

\end{description}

\subsection{Construction Principles}\label{sec:constrprinc}

There are different ways to construct anonymous credentials on a technical level. At a high level, the user obtains a signature from the issuer on a set of attributes, which typically includes some secret of the user that is not revealed to the issuer. For presentation, the user then demonstrates possession of such a valid signature from the issuer on a set of attributes that satisfies a certain relation. This can be just revealing a subset of the attributes or be some more complex relation. The important point is that the presentation does not reveal the original signature from the issuer directly and thus cannot be linked to the issuing. Also, what is revealed during the presentations is unlinkable. 
~\\[0.5em]\textbf{Zero-knowledge proofs.} If one wants to demonstrate a more complex relation, this typically requires the use of zero-knowledge proofs (ZKPs). We recall that ZKPs~\cite{DBLP:journals/siamcomp/GoldwasserMR89} allow one party (the prover) by interacting with another party (the verifier) to convince the latter that a statement (from some NP language) is true without revealing any additional information (the zero-knowledge property). At the same time, the prover is not able to make the verifier accept proofs about false statements (the soundness property). Often it is important to remove interaction in that the prover only needs to compute a single message (a proof), which can then be verified by everyone. These are so-called non-interactive zero-knowledge (NIZK) proofs \cite{DBLP:conf/stoc/BlumFM88}. It should be noted that deploying ZKPs or NIZKs introduces additional computational and implementation complexities and requires a certain level of cryptographic know-how.
~\\[0.5em]\textbf{Dedicated constructions.} Anonymous credentials with a focus on obtaining highly efficient instantiations can be constructed from specific types of signatures schemes. They can be based on signatures with efficient protocols like CL \cite{DBLP:conf/crypto/CamenischL04}, BBS \cite{DBLP:conf/crypto/BonehBS04,DBLP:conf/eurocrypt/TessaroZ23a}, BBS+ \cite{DBLP:conf/scn/AuSM06,DBLP:conf/trust/CamenischDL16} or PS \cite{DBLP:conf/ctrsa/PointchevalS16} signatures, which are specific signature schemes that are compatible with commitment schemes and efficient ZKPs. Alternatively, they can be based on signatures with specific randomization properties which allow to avoid ZKPs \cite{DBLP:conf/asiacrypt/Verheul01,DBLP:journals/joc/FuchsbauerHS19}, or on specific redactable signatures that allow to redact signed messages \cite{DBLP:conf/asiacrypt/CamenischDHK15,DBLP:conf/pkc/Sanders20}. 

As argued in the recent discussion in \cite{cryptoeudi}, BBS signatures are currently a popular choice for recent and ongoing projects. We will not take a closer look into how they or related schemes are constructed and for our purposes it suffices to consider all of them under the term \emph{dedicated constructions}.
~\\[0.5em]\textbf{Generic constructions.} A second and more generic way is to construct them from \emph{any} signature scheme, e.g., standardized and widely deployed ones like the Elliptic Curve Digital Signature Algorithm (ECDSA), and a NIZK proof system. While generically this does not yield concretely efficient (and actually rather theoretic) constructions, in the recent decade there has been enormous progress in the field of NIZK proofs and in particular on zero-knowledge succinct non-interactive arguments of knowledge (zk-SNARKs) (cf. \cite{Nitulescu2020zkSNARKsAG,SEC-030}). zk-SNARKs are succinct NIZK proofs, where loosely speaking succinctness means that the proofs are short and the verification of proofs is efficient. Using recent advances in this directions makes the construction of such anonymous credential schemes relying on existing signatures such as ECDSA practically feasible. There are very recent works by Google \cite{cryptoeprint:2024/2010} and Microsoft \cite{cryptoeprint:2024/2013} leveraging zk-SNARKs to turn existing credentials (i.e., ECDSA and ECDSA or RSA signatures respectively) into anonymous credentials.

\section{Anonymous Credentials: The Practice}\label{sec:anonpractice}
{}

It is fair to say that when considering industry adoption, anonymous credentials have not been a success story so far. Fortunately, the picture is changing and we see increasing interest in deploying such technology. In this section we want to look at the anonymous credential technology from an implementation, standardization and deployment perspective.

\subsection{Implementations}
To the best of our knowledge, the first implementation of a multi-use anonymous credential system was the idemix system developed within IBM in the early 2000's \cite{DBLP:conf/ccs/CamenischH02} (using strong RSA-based CL signatures \cite{DBLP:conf/scn/CamenischL02}) and further developed and improved among others over a series of EU projects such as the EU FP7 projects PrimeLife and ABC4Trust. Later, the idemix project has been integrated into the Hyperledger project, now sponsored and managed by the Linux Foundation\footnote{\url{https://www.linuxfoundation.org}}, and is the basis for the Hyperledger AnonCreds\footnote{\url{https://lf-hyperledger.atlassian.net/wiki/spaces/ANONCREDS/overview}} and related projects (e.g., Aries Bifold or Hyperledger Aries), which plan to move to BBS signatures in its version 2.0 \cite{BeKaBBS23}. Moreover, the first implementation of a single-use anonymous credential system relying on the techniques in \cite{BrandsBook} was due to Brands at Credentica in the early 2000's and was then acquired by and continued as the U-Prove project at Microsoft\footnote{\url{https://www.microsoft.com/en-us/research/project/u-prove/}}.

We note that in the field of distributed ledger technology and verifiable credentials, there are meanwhile numerous open-source implementations of anonymous credentials based on BBS(+) signatures. We may mention here the Rust bbs\footnote{\url{https://crates.io/crates/bbs}} or ursa crate\footnote{\url{https://docs.rs/ursa/}}, the docknetwork implementation\footnote{\url{https://github.com/docknetwork/crypto}} and the BBS implementation of Digital Bazaar\footnote{\url{https://github.com/digitalbazaar/bbs-signatures}}, provided solely as some illustrative instances. We also refer the interested reader to a recent work by Flamini et al. \cite{DBLP:journals/istr/FlaminiSSSTR24} for a comparison and experimental evaluation.

The anonymous credentials zoo\footnote{\url{https://tokenzoo.github.io}} provides an (somewhat outdated) overview of some multi- as well as single-use anonymous credentials with some pointers to implementations. Unfortunately, we are not aware of a place that compiles a list of current open-source implementations, something that would be very helpful and highly desirable.

\subsection{Standardization Efforts}

There are various ongoing efforts when it comes to the standardization of anonymous credentials and related technology. It is important to note that standardization is of utmost importance to guarantee interoperability of different implementations and thus the compatibility of services and products that rely on this technology.

First, we want to mention the Crypto Forum Research Group (CFRG) within the Internet Engineering Task Force (IETF), an open, global engineering community that publishes de facto standards called Requests For Comments (RFCs). There are ongoing standardization processes for BBS signatures \cite{I-D.irtf-cfrg-bbs-signatures}, blind BBS signatures \cite{I-D.kalos-bbs-blind-signatures} as well as pseudonyms for BBS signatures \cite{I-D.kalos-bbs-per-verifier-linkability}. Moreover, there are many (ongoing) standardization efforts for related primitives such as oblivious pseudo-random functions (OPRFs) or blind signatures.

In addition, there are non-profit organizations such as the aforementioned Linux Foundation or industry consortia such as the Decentralized Identity Foundation (DIF)\footnote{\url{https://identity.foundation}}, the World Wide Web Consortium (W3C)\footnote{\url{https://www.w3.org}}  or ZKProof\footnote{\url{https://zkproof.org}} which are running working groups on anonymous credential technologies and the underlying technologies respectively. 

Finally, there is the International Organization for Standardization (ISO) currently running a preliminary work item (ISO/IEC PWI 24843) on Privacy-Preserving Attribute-Based Credentials within the ISO/IEC JTC1/SC 27 WG 2 and on guidelines on privacy preservation based on zero-knowledge proofs (ISO/IEC CD 27565). Moreover, the European Telecommunications Standards Institute (ETSI) has several technical reports on related technology, such as ETSI TR 119 476\footnote{\href{https://www.etsi.org/deliver/etsi_tr/119400_119499/119476/01.02.01_60/tr_119476v010201p.pdf}{\tt https://www.etsi.org/deliver/etsi\textunderscore tr/119476}} on selective disclosure and zero-knowledge proofs.

\subsection{Real-World Deployments}

The biggest real-world use cases related to anonymous credential technology and rolled out on billions of devices are the Direct Anonymous Attestation (DAA) scheme implemented in any Trusted Platform Module (TPM) or the Enhanced Privacy ID (EPID)\footnote{\href{https://www.intel.com/content/www/us/en/developer/articles/technical/intel-enhanced-privacy-id-epid-security-technology.html}{\tt https://www.intel.com/intel-enhanced-privacy-id-epid-security-technology}} available in Intel's SGX. However, while somewhat close to multi-use anonymous credentials, those schemes were designed for specific use cases, do not consider attributes, and are thus not generally applicable.

When it comes to multi-use anonymous credentials in the conventional sense, we want to mention the Yivi App (previously known as IRMA) for smartphone use in the Netherlands\footnote{\url{https://privacybydesign.foundation/irma-explanation/}}, which is based upon idemix. Finally, privately verifiable multi-use anonymous credentials have been used to implement the feature of private groups in the Signal messenger\footnote{\url{https://signal.org/blog/signal-private-group-system/}}.

There are also various governmental use cases that rely on multi-use anonymous credentials, e.g., supporting sustainable economic development by the government of British Columbia (Canada) or digital travel credentials by the government of Aruba.  We refer to \cite{cryptoeudi} for more discussion on such use cases.

Apart from this, we want to mention the use of so-called anonymous tokens in various forms and use cases. Those represent single-use credentials, with or without attributes that are either privately or publicly available. This includes Privacy Pass \cite{DBLP:journals/popets/DavidsonGSTV18} as used by Cloudflare\footnote{\url{https://blog.cloudflare.com/privacy-pass-standard/}}, or the Anonymous Credential Service by Meta\footnote{\url{https://engineering.fb.com/2022/12/12/security/anonymous-credential-service-acs-open-source/}}.

\section{European Union Digital Identity Wallet (EUDIW)}\label{sec:euidw}

With the EU Digital Identity Framework Regulation\footnote{\url{https://oeil.secure.europarl.europa.eu/oeil/en/procedure-file?reference=2021/0136(COD)}} (commonly known as eIDAS 2.0) entering into force in May 2024, the EU proposes the EU Digital Identity Wallet (EUDIW), which shall be a “fully mobile, secure and user-friendly” service, enabling users to identify themselves
to public and private online services. Consequently, all member states are required to offer such an EUDIW to their citizens and residents by 2026. We refer to \cite{cryptoeudi} for a more thorough discussion, and will only briefly discuss the initial design presented in the architectural reference framework (ARF) version 1.4.0 \cite{arf} and its shortcomings from a privacy perspective.

\subsection{Signatures on Salted Hashes and their Privacy Issues}\label{sec:saltedhash}
The approach pursued in the ARF version 1.4.0 \cite{arf} is to follow the realization of selective attribute disclosure credentials from conventional digital signatures (e.g., ECDSA) as specified in the ISO/IEC 18013-5 (mobile driving license) standard. At a high level, the idea is that an issuer signs a list of ``salted hashes'' $(h_1,\ldots,h_n)$ instead of directly signing the attributes $(a_1,\ldots,a_n)$. Each salted hash $h_i=H(a_i||r_i)$ is computed using a cryptographic hash function $H$ such as SHA-3 on input the attribute $a_i$ concatenated with a sufficiently long uniformly random string $r_i$. This can be viewed as a hash commitment, i.e., $h_i$ hides $a_i$ as long as $r_i$ is not known and hash value $h_i$ cannot be opened to a different $a_i'\neq a_i$, and thus is binding. Consequently, the commitment hides the attribute until the salt is revealed; even if the same attribute appears in different credentials, the added salt ensures that each hash is unique, thereby preventing straightforward linkage.

The idea is now that when a user presents a signature for $(h_1,\ldots,h_n)$ to some RP for attributes it wants to present, it reveals $a_i$ along with $r_i$. For all attributes that are not disclosed, the user simply reveals $h_j$, which as discussed does not reveal anything about $a_j$ (as $r_j$ is not revealed). 
~\\[0.5em]\textbf{Privacy discussion.} This represents the functionality of selective disclosure, i.e., the user can decide which attributes $a_i$ to reveal. However, when presenting the same signature multiple times, the signature as well as $(h_1,\ldots,h_n)$ are always revealed and thus can be trivially linked by any RP. This can be avoided by presenting every signature only once, i.e., viewing them as single-use credentials, which comes at the cost of requiring frequent re-issuance of new credentials and keeping track of already used signatures on the user side, and may reduce convenience. Moreover, the collusion of the IdP and any RP always completely breaks unlinkability as the IdP issues all signatures and thereby sees all attributes. We also refer the interested reader to \cite{cryptoeudi} for a more thorough discussion of these aspects.  

\subsection{Privacy-Oriented and Desirable Solution}
Now having discussed multi-use anonymous credentials in Section \ref{sec:anontheory} and Section \ref{sec:anonpractice}, it seems natural to consider this tool as a solution to the privacy problems raised above. Obviously, all the variants of unlinkability can be satisfactorily addressed by using anonymous credentials. And indeed, this is what a number of cryptographers suggested when invited by the EUDI Wallet Team of the European Commission to provide feedback concerning attestations and zero-knowledge proofs with respect to the ARF version 1.4.0 \cite{arf}. More precisely, the expert group suggested the use of anonymous credentials based on BBS signatures, which can be instantiated using pairing-friendly elliptic curves \cite{cryptoeudi}. In the next section we will discuss why, although considered mature enough, such anonymous credentials cannot be readily deployed.

\section{Barriers to Widespread Adoption}\label{sec:barriers}

While there might be numerous reasons that slow down or prevent a widespread adoption of ``new'' technologies such as anonymous credentials, we will solely focus on technical aspects. And even here, we do not claim to be exhaustive and do not cover important aspects such as ensuring end-user usability or system interoperability. We will discuss two aspects which we consider really important (and non-trivial to solve) issues that complicate practical adoption. They have also shown up in a recent discussion about the EUDIW after providing the feedback in \cite{cryptoeudi}. Moreover, at the end of this section, we will also briefly discuss challenges that are arising from the recently started transitions to post-quantum cryptography.

\subsection{Dependency on Established Standards}
If the decision of deploying anonymous credentials depends on the availability of standards for AC schemes, then this can result in significant delays. Having no standards available means that even if standardization is to be started immediately, depending on the standardization body, the time until a standard is finalized might take several years. This is especially true for large international standardization bodies like ISO or ETSI. For de-facto standards like the ones by the Internet Engineering Task Force (IETF) or the World Wide Web Consortium (W3C) this can be significantly shorter, but still typically requires several months to years. Nevertheless, in the latter case there are public versions available from the first draft, so that adoption can potentially start earlier. 

Moreover, there might be standards that do not directly relate to the AC schemes and their technical details, that still can influence potential deployments of such schemes. One example is how credentials in such schemes are specified when it comes to the document structure. Here, the two currently dominating standards are verifiable credentials by the W3C\footnote{\url{https://www.w3.org/TR/vc-overview/}} and the mdoc format specified in the ISO/IEC 18013 family of documents that specify the technical and operational requirements for physical and mobile driver's licenses (mDL). While in the design of the former the use of anonymous credentials has been considered, the latter does not do so. Actually, it is considered along with a dedicated solution to selective disclosure, which as we have discussed in Section \ref{sec:saltedhash} does not provide a sufficient degree of unlinkability. This might introduce overhead and the requirement for ad-hoc intermediate formats when using them along with anonymous credentials.

\subsection{Hardware Support and Compatibility}\label{sec:hwcomp}
While only requiring pure software implementations of new technologies such as anonymous credentials is a manageable task, the integration of new cryptographic approaches into existing ecosystems and infrastructure is a much more delicate task. In the former case it boils down to secure implementations of cryptography, which by itself is a non-trivial task and requires lots of expert knowledge. But it does not suffer from critical external dependencies. However, in the latter case it requires backward compatibility at many places. For instance, cryptographic keys and operations involving secret keys (such as signing) are typically handled by dedicated hardware such as hardware security modules (HSM), Trusted Platform Modules (TPMs), or other secure elements such as smart cards. Alternatively, they can be handled by Trusted Execution Environments (TEEs) such as ARM's TrustZone or Intel's SGX. Irrespective of the concrete realization, what is common to all those technologies is that the provided cryptographic functionality is limited to very few cryptographic primitives. When it comes to signature functionality, this typically either means RSA or ECDSA, and, in the latter case, specific standardized elliptic curves such as curve P-256 (also known as secp256r1 or prime256v1).

However, anonymous credentials such as those based on BBS \cite{DBLP:conf/crypto/BonehBS04}, as suggested in \cite{cryptoeudi}, require specific pairing-friendly elliptic curves such as BN-462 or BLS12-381\footnote{\url{ https://members.loria.fr/AGuillevic/pairing-friendly-curves/}} that are not readily available in the aforementioned technologies. Moreover, this requires other interfaces as those available for conventional signatures. Consequently, such a functionality is simply not available in currently deployed secure elements and TEEs. The only functionality that comes close is that of Direct Anonymous Attestation (DAA) in TPMs or EPID available in Intel's SGX. However, as already mentioned they were designed for the specific use cases and do not consider attributes in any way. While making anonymous credentials available in TEEs seems easier than when they are rooted in hardware, such changes might also require significant changes to software stacks, e.g., client platforms such as the Android system. Especially as such systems also require backward compatibility. A very informative recent talk on changing legacy cryptography and the associated problems\footnote{\url{ https://youtu.be/-YXCojP8IjE}} by abhi shelat is highly recommended at this place.

\subsection{Transitioning to Post-Quantum Cryptography}
In response to the threats by scalable quantum computers to traditional public-key cryptography, the first standards for post-quantum cryptographic schemes are meanwhile available from the National Institute of Standards and Technology (NIST)\footnote{\url{https://www.nist.gov/pqcrypto}}. Many governments are currently pushing towards post-quantum and are developing roadmaps for the transition from traditional to post-quantum cryptography. Clearly, this topic is  also important for anonymous credential schemes. Consequently, it is strongly advisable to consider the post-quantum aspect in any current design so that a switch to a quantum safe replacement for the used anonymous credential scheme at some point does not require a partial or even complete re-design. In other words, it is important that a solution that is being deployed from now onward provides cryptographic agility. 

It is important to mention though, that while we now see the first NIST standards for basic public-key cryptographic primitives (i.e., public-key encryption and signatures), we are still not close to having a clear picture of candidates for post-quantum anonymous credentials. There are recent constructions of lattice-based anonymous credentials \cite{DBLP:conf/crypto/BootleLNS23,DBLP:journals/iacr/ArgoGJLRS24,cryptoeprint:2025/356} and first proof-of-concept implementations are available from IBM \cite{cryptoeprint:2024/1846} or the EU QUBIP project\footnote{\url{https://github.com/Cybersecurity-LINKS/pqzk-blns}}. While these constructions do not yet seem well-studied enough in terms of security and efficiency, it can, however, be expected that significant research will go into this topic in the upcoming years. Moreover, as discussed in more detail in \cite{cryptoeudi}, post-quantum security is less critical in authentication primitives than it is for encryption. Also it should be noted that if the privacy property of an anonymous credential system is unconditional (as it is the case for the BBS family mentioned in Section \ref{sec:constrprinc}), then even when instantiated from building blocks that could be broken by a hypothetical quantum computer, the privacy is not endangered and will hold forever (even when given unlimited computing power).

\section{Conclusions and Outlook}\label{sec:conclusion}

It needs to be concluded that currently enrolled (and partly also planned) identity solutions typically provide rather weak privacy protection and are quite far from what could be considered the ideal case. Anonymous credentials are the right tool to achieve strong privacy protection and are well studied and well understood by the research community. However, although they have already been proposed back in the early 1980s by Chaum, and a large body of literature as well as practical use cases and open-source implementations exist, for a long time they did not see adoption in industry. Fortunately, in recent years we see a growing interest from the industry and governments to adopt such privacy-preserving technologies. However, deploying new technology comes with huge (financial) efforts, particularly when use cases require backward compatibility with existing hardware that only implements ``legacy cryptography'', which typically cannot be readily used with approaches and in particular modern anonymous credentials. Recent progress in cryptographic research and in particular in the field of zk-SNARKs, however, does now even enable solutions based on existing ``legacy cryptography'' and so even immediate large-scale deployments do not seem out of reach. 

Nevertheless, when deploying solutions now, they are likely here to stay for quite some time. Thus, solid and forward-looking planning is essential. This particularly holds true for the governmental domain and large scale solutions like the EUDIW. For instance, when considering the rollout a new solution today, it is definitely advisable to consider quantum resistance (i.e., post-quantum security). 

At the time of writing this paper, the technical implementation of the EUDIW is still under discussion.\footnote{\url{https://github.com/eu-digital-identity-wallet/eudi-doc-architecture-and-reference-framework/discussions}} At this point it cannot be determined with certainty how the final version of the ARF will look like and which technology will ultimately be adopted. We strongly hope that it will be a solution that considers strong privacy protection.   
~\\[0.5em]\textbf{Acknowledgments.} The author is very grateful to Yod Samuel Mart{\'{\i}}n, Ren{\'{e}} Mayrhofer, Omid Mir, Octavio Perez{-}Kempner and Mahdi Sedaghat for their helpful feedback on a draft of this paper.



%
%
%
%
\ifsubmission
\bibliographystyle{splncs04}
\bibliography{refs}

@misc{cryptoeprint:2025/356,
      author = {Adrien Dubois and Michael Klooß and Russell W. F. Lai and Ivy K. Y. Woo},
      title = {Lattice-based Proof-Friendly Signatures from Vanishing Short Integer Solutions},
      howpublished = {Cryptology {ePrint} Archive, Paper 2025/356},
      year = {2025},
      url = {https://eprint.iacr.org/2025/356}
}

@inproceedings{DBLP:conf/ndss/Garman0M14,
  author       = {Christina Garman and
                  Matthew Green and
                  Ian Miers},
  title        = {Decentralized Anonymous Credentials},
  booktitle    = {21st Annual Network and Distributed System Security Symposium, {NDSS}
                  2014, San Diego, California, USA, February 23-26, 2014},
  publisher    = {The Internet Society},
  year         = {2014},
  url          = {https://www.ndss-symposium.org/ndss2014/decentralized-anonymous-credentials},
  bibsource    = {dblp computer science bibliography, https://dblp.org}
}

@inproceedings{DBLP:conf/sp/RosenbergWGM23,
  author       = {Michael Rosenberg and
                  Jacob D. White and
                  Christina Garman and
                  Ian Miers},
  title        = {zk-creds: Flexible Anonymous Credentials from zkSNARKs and Existing
                  Identity Infrastructure},
  booktitle    = {44th {IEEE} Symposium on Security and Privacy, {SP} 2023, San Francisco,
                  CA, USA, May 21-25, 2023},
  pages        = {790--808},
  publisher    = {{IEEE}},
  year         = {2023},
  url          = {https://doi.org/10.1109/SP46215.2023.10179430},
  doi          = {10.1109/SP46215.2023.10179430},
  bibsource    = {dblp computer science bibliography, https://dblp.org}
}

@phdthesis{sso-pashalidis,
  title        = {Interdomain User Authentication and Privacy},
  author       = {Andreas Pashalidis},
  year         = 2006,
  note         = {\url{https://web.archive.org/web/20060925104053/http://www.ma.rhul.ac.uk/techreports/2005/RHUL-MA-2005-13.pd}},
  school       = {Royal Holloway, University of London},
  type         = {PhD thesis}
}

@article{DBLP:journals/istr/FlaminiSSSTR24,
  author       = {Andrea Flamini and
                  Giada Sciarretta and
                  Mario Scuro and
                  Amir Sharif and
                  Alessandro Tomasi and
                  Silvio Ranise},
  title        = {On cryptographic mechanisms for the selective disclosure of verifiable
                  credentials},
  journal      = {J. Inf. Secur. Appl.},
  volume       = {83},
  pages        = {103789},
  year         = {2024},
  url          = {https://doi.org/10.1016/j.jisa.2024.103789},
  doi          = {10.1016/J.JISA.2024.103789},
  bibsource    = {dblp computer science bibliography, https://dblp.org}
}

@inproceedings{DBLP:conf/sp/DoernerKLST23,
  author       = {Jack Doerner and
                  Yashvanth Kondi and
                  Eysa Lee and
                  Abhi Shelat and
                  LaKyah Tyner},
  title        = {Threshold {BBS+} Signatures for Distributed Anonymous Credential Issuance},
  booktitle    = {44th {IEEE} Symposium on Security and Privacy, {SP} 2023, San Francisco,
                  CA, USA, May 21-25, 2023},
  pages        = {773--789},
  publisher    = {{IEEE}},
  year         = {2023},
  url          = {https://doi.org/10.1109/SP46215.2023.10179470},
  doi          = {10.1109/SP46215.2023.10179470},
  bibsource    = {dblp computer science bibliography, https://dblp.org}
}

@article{DBLP:journals/iacr/Orru24,
  author       = {Michele Orr{\`{u}}},
  title        = {Revisiting Keyed-Verification Anonymous Credentials},
  journal      = {{IACR} Cryptol. ePrint Arch.},
  pages        = {1552},
  year         = {2024},
  url          = {https://eprint.iacr.org/2024/1552},
  bibsource    = {dblp computer science bibliography, https://dblp.org}
}

@inproceedings{DBLP:conf/eurocrypt/TessaroZ23a,
  author       = {Stefano Tessaro and
                  Chenzhi Zhu},
  editor       = {Carmit Hazay and
                  Martijn Stam},
  title        = {Revisiting {BBS} Signatures},
  booktitle    = {Advances in Cryptology - {EUROCRYPT} 2023 - 42nd Annual International
                  Conference on the Theory and Applications of Cryptographic Techniques,
                  Lyon, France, April 23-27, 2023, Proceedings, Part {V}},
  series       = {Lecture Notes in Computer Science},
  volume       = {14008},
  pages        = {691--721},
  publisher    = {Springer},
  year         = {2023},
  url          = {https://doi.org/10.1007/978-3-031-30589-4_24},
  doi          = {10.1007/978-3-031-30589-4_24},
  bibsource    = {dblp computer science bibliography, https://dblp.org}
}

@inproceedings{DBLP:conf/acns/BaldimtsiCHKLN15,
  author       = {Foteini Baldimtsi and
                  Jan Camenisch and
                  Lucjan Hanzlik and
                  Stephan Krenn and
                  Anja Lehmann and
                  Gregory Neven},
  editor       = {Tal Malkin and
                  Vladimir Kolesnikov and
                  Allison Bishop Lewko and
                  Michalis Polychronakis},
  title        = {Recovering Lost Device-Bound Credentials},
  booktitle    = {Applied Cryptography and Network Security - 13th International Conference,
                  {ACNS} 2015, New York, NY, USA, June 2-5, 2015, Revised Selected Papers},
  series       = {Lecture Notes in Computer Science},
  volume       = {9092},
  pages        = {307--327},
  publisher    = {Springer},
  year         = {2015},
  url          = {https://doi.org/10.1007/978-3-319-28166-7_15},
  doi          = {10.1007/978-3-319-28166-7_15},
  bibsource    = {dblp computer science bibliography, https://dblp.org}
}

@misc{cryptoeprint:2024/1874,
      author = {Andrea Flamini and Eysa Lee and Anna Lysyanskaya},
      title = {Multi-Holder Anonymous Credentials from {BBS} Signatures},
      howpublished = {Cryptology {ePrint} Archive, Paper 2024/1874},
      url = {https://eprint.iacr.org/2024/1874},
      year = {2024}
}

@inproceedings{DBLP:conf/uss/YeohKHKH23,
  author       = {Wei{-}Zhu Yeoh and
                  Michal Kepkowski and
                  Gunnar Heide and
                  Dali Kaafar and
                  Lucjan Hanzlik},
  editor       = {Joseph A. Calandrino and
                  Carmela Troncoso},
  title        = {Fast IDentity Online with Anonymous Credentials {(FIDO-AC)}},
  booktitle    = {32nd {USENIX} Security Symposium, {USENIX} Security 2023, Anaheim,
                  CA, USA, August 9-11, 2023},
  pages        = {3029--3046},
  publisher    = {{USENIX} Association},
  year         = {2023},
  url          = {https://www.usenix.org/conference/usenixsecurity23/presentation/yeoh},
  bibsource    = {dblp computer science bibliography, https://dblp.org}
}

@article{DBLP:journals/popets/KroschewskiL23,
  author       = {Maximilian Kroschewski and
                  Anja Lehmann},
  title        = {Save The Implicit Flow? Enabling Privacy-Preserving {RP} Authentication
                  in OpenID Connect},
  journal      = {Proc. Priv. Enhancing Technol.},
  volume       = {2023},
  number       = {4},
  pages        = {96--116},
  year         = {2023},
  url          = {https://doi.org/10.56553/popets-2023-0100},
  doi          = {10.56553/POPETS-2023-0100},
  bibsource    = {dblp computer science bibliography, https://dblp.org}
}

@misc{cryptoeprint:2024/1124,
      author = {Maximilian Kroschewski and Anja Lehmann and Cavit {\"O}zbay},
      title = {{OPPID}: Single Sign-On with Oblivious Pairwise Pseudonyms},
      howpublished = {Cryptology {ePrint} Archive, Paper 2024/1124},
      year = {2024},
      url = {https://eprint.iacr.org/2024/1124}
}

@inproceedings{DBLP:conf/soups/WashRBW16,
  author       = {Rick Wash and
                  Emilee J. Rader and
                  Ruthie Berman and
                  Zac Wellmer},
  title        = {Understanding Password Choices: How Frequently Entered Passwords Are
                  Re-used across Websites},
  booktitle    = {Twelfth Symposium on Usable Privacy and Security, {SOUPS} 2016, Denver,
                  CO, USA, June 22-24, 2016},
  pages        = {175--188},
  publisher    = {{USENIX} Association},
  year         = {2016},
  url          = {https://www.usenix.org/conference/soups2016/technical-sessions/presentation/wash},
  bibsource    = {dblp computer science bibliography, https://dblp.org}
}

@inproceedings{DBLP:conf/soups/ZhangPBC19,
  author       = {Shikun Zhang and
                  Sarah Pearman and
                  Lujo Bauer and
                  Nicolas Christin},
  editor       = {Heather Richter Lipford},
  title        = {Why people (don't) use password managers effectively},
  booktitle    = {Fifteenth Symposium on Usable Privacy and Security, {SOUPS} 2019,
                  Santa Clara, CA, USA, August 11-13, 2019},
  publisher    = {{USENIX} Association},
  year         = {2019},
  url          = {https://www.usenix.org/conference/soups2019/presentation/pearman},
  bibsource    = {dblp computer science bibliography, https://dblp.org}
}

@misc{cryptoeprint:2023/1039,
      author = {Alishah Chator and Matthew Green and Pratyush Ranjan Tiwari},
      title = {{SoK}: Privacy-Preserving Signatures},
      howpublished = {Cryptology {ePrint} Archive, Paper 2023/1039},
      year = {2023},
      url = {https://eprint.iacr.org/2023/1039}
}

@inproceedings{DBLP:conf/secsr/KakviMPQ23,
  author       = {Saqib A. Kakvi and
                  Keith M. Martin and
                  Colin Putman and
                  Elizabeth A. Quaglia},
  editor       = {Felix G{\"{u}}nther and
                  Julia Hesse},
  title        = {SoK: Anonymous Credentials},
  booktitle    = {Security Standardisation Research - 8th International Conference,
                  {SSR} 2023, Lyon, France, April 22-23, 2023, Proceedings},
  series       = {Lecture Notes in Computer Science},
  volume       = {13895},
  pages        = {129--151},
  publisher    = {Springer},
  year         = {2023},
  url          = {https://doi.org/10.1007/978-3-031-30731-7_6},
  doi          = {10.1007/978-3-031-30731-7_6},
  bibsource    = {dblp computer science bibliography, https://dblp.org}
}

@book{BrandsBook,
author = {Brands, Stefan A.},
title = {Rethinking Public Key Infrastructures and Digital Certificates: Building in Privacy},
year = {2000},
isbn = {0262024918},
publisher = {MIT Press},
address = {Cambridge, MA, USA}
}

@inproceedings{DBLP:conf/scn/AuSM06,
  author       = {Man Ho Au and
                  Willy Susilo and
                  Yi Mu},
  editor       = {Roberto De Prisco and
                  Moti Yung},
  title        = {Constant-Size Dynamic \emph{k}-TAA},
  booktitle    = {Security and Cryptography for Networks, 5th International Conference,
                  {SCN} 2006, Maiori, Italy, September 6-8, 2006, Proceedings},
  series       = {Lecture Notes in Computer Science},
  volume       = {4116},
  pages        = {111--125},
  publisher    = {Springer},
  year         = {2006},
  url          = {https://doi.org/10.1007/11832072_8},
  doi          = {10.1007/11832072_8},
  bibsource    = {dblp computer science bibliography, https://dblp.org}
}

@inproceedings{DBLP:conf/trust/CamenischDL16,
  author       = {Jan Camenisch and
                  Manu Drijvers and
                  Anja Lehmann},
  editor       = {Michael Franz and
                  Panos Papadimitratos},
  title        = {Anonymous Attestation Using the Strong Diffie Hellman Assumption Revisited},
  booktitle    = {Trust and Trustworthy Computing - 9th International Conference, {TRUST}
                  2016, Vienna, Austria, August 29-30, 2016, Proceedings},
  series       = {Lecture Notes in Computer Science},
  volume       = {9824},
  pages        = {1--20},
  publisher    = {Springer},
  year         = {2016},
  url          = {https://doi.org/10.1007/978-3-319-45572-3_1},
  doi          = {10.1007/978-3-319-45572-3_1},
  bibsource    = {dblp computer science bibliography, https://dblp.org}
}

@inproceedings{DBLP:conf/eurocrypt/CamenischL01,
  author       = {Jan Camenisch and
                  Anna Lysyanskaya},
  editor       = {Birgit Pfitzmann},
  title        = {An Efficient System for Non-transferable Anonymous Credentials with
                  Optional Anonymity Revocation},
  booktitle    = {Advances in Cryptology - {EUROCRYPT} 2001, International Conference
                  on the Theory and Application of Cryptographic Techniques, Innsbruck,
                  Austria, May 6-10, 2001, Proceeding},
  series       = {Lecture Notes in Computer Science},
  volume       = {2045},
  pages        = {93--118},
  publisher    = {Springer},
  year         = {2001},
  url          = {https://doi.org/10.1007/3-540-44987-6_7},
  doi          = {10.1007/3-540-44987-6_7},
  bibsource    = {dblp computer science bibliography, https://dblp.org}
}

@article{DBLP:journals/joc/FuchsbauerHS19,
  author       = {Georg Fuchsbauer and
                  Christian Hanser and
                  Daniel Slamanig},
  title        = {Structure-Preserving Signatures on Equivalence Classes and Constant-Size
                  Anonymous Credentials},
  journal      = {J. Cryptol.},
  volume       = {32},
  number       = {2},
  pages        = {498--546},
  year         = {2019},
  url          = {https://doi.org/10.1007/s00145-018-9281-4},
  doi          = {10.1007/S00145-018-9281-4},
  bibsource    = {dblp computer science bibliography, https://dblp.org}
}

@inproceedings{DBLP:conf/ndss/SonninoABMD19,
  author       = {Alberto Sonnino and
                  Mustafa Al{-}Bassam and
                  Shehar Bano and
                  Sarah Meiklejohn and
                  George Danezis},
  title        = {Coconut: Threshold Issuance Selective Disclosure Credentials with
                  Applications to Distributed Ledgers},
  booktitle    = {26th Annual Network and Distributed System Security Symposium, {NDSS}
                  2019, San Diego, California, USA, February 24-27, 2019},
  publisher    = {The Internet Society},
  year         = {2019},
  bibsource    = {dblp computer science bibliography, https://dblp.org},
  url= {https://www.ndss-symposium.org/ndss-paper/coconut-threshold-issuance-selective-disclosure-credentials-with-applications-to-distributed-ledgers/}
}

@inproceedings{DBLP:conf/asiacrypt/Verheul01,
  author       = {Eric R. Verheul},
  editor       = {Colin Boyd},
  title        = {Self-Blindable Credential Certificates from the Weil Pairing},
  booktitle    = {Advances in Cryptology - {ASIACRYPT} 2001, 7th International Conference
                  on the Theory and Application of Cryptology and Information Security,
                  Gold Coast, Australia, December 9-13, 2001, Proceedings},
  series       = {Lecture Notes in Computer Science},
  volume       = {2248},
  pages        = {533--551},
  publisher    = {Springer},
  year         = {2001},
  url          = {https://doi.org/10.1007/3-540-45682-1_31},
  doi          = {10.1007/3-540-45682-1_31},
  bibsource    = {dblp computer science bibliography, https://dblp.org}
}

@article{DBLP:journals/siamcomp/GoldwasserMR89,
  author       = {Shafi Goldwasser and
                  Silvio Micali and
                  Charles Rackoff},
  title        = {The Knowledge Complexity of Interactive Proof Systems},
  journal      = {{SIAM} J. Comput.},
  volume       = {18},
  number       = {1},
  pages        = {186--208},
  year         = {1989},
  url          = {https://doi.org/10.1137/0218012},
  doi          = {10.1137/0218012},
  bibsource    = {dblp computer science bibliography, https://dblp.org}
}

@inproceedings{DBLP:conf/pkc/Sanders20,
  author       = {Olivier Sanders},
  editor       = {Aggelos Kiayias and
                  Markulf Kohlweiss and
                  Petros Wallden and
                  Vassilis Zikas},
  title        = {Efficient Redactable Signature and Application to Anonymous Credentials},
  booktitle    = {Public-Key Cryptography - {PKC} 2020 - 23rd {IACR} International Conference
                  on Practice and Theory of Public-Key Cryptography, Edinburgh, UK,
                  May 4-7, 2020, Proceedings, Part {II}},
  series       = {Lecture Notes in Computer Science},
  volume       = {12111},
  pages        = {628--656},
  publisher    = {Springer},
  year         = {2020},
  url          = {https://doi.org/10.1007/978-3-030-45388-6_22},
  doi          = {10.1007/978-3-030-45388-6_22},
  bibsource    = {dblp computer science bibliography, https://dblp.org}
}

@inproceedings{DBLP:conf/asiacrypt/CamenischDHK15,
  author       = {Jan Camenisch and
                  Maria Dubovitskaya and
                  Kristiyan Haralambiev and
                  Markulf Kohlweiss},
  editor       = {Tetsu Iwata and
                  Jung Hee Cheon},
  title        = {Composable and Modular Anonymous Credentials: Definitions and Practical
                  Constructions},
  booktitle    = {Advances in Cryptology - {ASIACRYPT} 2015 - 21st International Conference
                  on the Theory and Application of Cryptology and Information Security,
                  Auckland, New Zealand, November 29 - December 3, 2015, Proceedings,
                  Part {II}},
  series       = {Lecture Notes in Computer Science},
  volume       = {9453},
  pages        = {262--288},
  publisher    = {Springer},
  year         = {2015},
  url          = {https://doi.org/10.1007/978-3-662-48800-3_11},
  doi          = {10.1007/978-3-662-48800-3_11},
  bibsource    = {dblp computer science bibliography, https://dblp.org}
}

@inproceedings{DBLP:conf/ctrsa/PointchevalS16,
  author       = {David Pointcheval and
                  Olivier Sanders},
  editor       = {Kazue Sako},
  title        = {Short Randomizable Signatures},
  booktitle    = {Topics in Cryptology - {CT-RSA} 2016 - The Cryptographers' Track at
                  the {RSA} Conference 2016, San Francisco, CA, USA, February 29 - March
                  4, 2016, Proceedings},
  series       = {Lecture Notes in Computer Science},
  volume       = {9610},
  pages        = {111--126},
  publisher    = {Springer},
  year         = {2016},
  url          = {https://doi.org/10.1007/978-3-319-29485-8_7},
  doi          = {10.1007/978-3-319-29485-8_7},
  bibsource    = {dblp computer science bibliography, https://dblp.org}
}

@inproceedings{DBLP:conf/ccs/CamenischH02,
  author       = {Jan Camenisch and
                  Els Van Herreweghen},
  editor       = {Vijayalakshmi Atluri},
  title        = {Design and implementation of the \emph{idemix} anonymous credential
                  system},
  booktitle    = {Proceedings of the 9th {ACM} Conference on Computer and Communications
                  Security, {CCS} 2002, Washington, DC, USA, November 18-22, 2002},
  pages        = {21--30},
  publisher    = {{ACM}},
  year         = {2002},
  url          = {https://doi.org/10.1145/586110.586114},
  doi          = {10.1145/586110.586114},
  bibsource    = {dblp computer science bibliography, https://dblp.org}
}

@misc{arf,
author={ARF},
 title = {The european digital identity wallet architecture and reference framework version 1.4.0},
 howpublished = {\url{https://eu-digital-identity-wallet.github.io/eudi-doc-architecture-and-reference-framework/1.4.0/arf/}},
 year = {2024}
}

@inproceedings{DBLP:conf/crypto/BonehBS04,
  author       = {Dan Boneh and
                  Xavier Boyen and
                  Hovav Shacham},
  editor       = {Matthew K. Franklin},
  title        = {Short Group Signatures},
  booktitle    = {Advances in Cryptology - {CRYPTO} 2004, 24th Annual International
                  CryptologyConference, Santa Barbara, California, USA, August 15-19,
                  2004, Proceedings},
  series       = {Lecture Notes in Computer Science},
  volume       = {3152},
  pages        = {41--55},
  publisher    = {Springer},
  year         = {2004},
  url          = {https://doi.org/10.1007/978-3-540-28628-8_3},
  doi          = {10.1007/978-3-540-28628-8_3},
  bibsource    = {dblp computer science bibliography, https://dblp.org}
}

@inproceedings{DBLP:conf/scn/CamenischL02,
  author       = {Jan Camenisch and
                  Anna Lysyanskaya},
  editor       = {Stelvio Cimato and
                  Clemente Galdi and
                  Giuseppe Persiano},
  title        = {A Signature Scheme with Efficient Protocols},
  booktitle    = {Security in Communication Networks, Third International Conference,
                  {SCN} 2002, Amalfi, Italy, September 11-13, 2002. Revised Papers},
  series       = {Lecture Notes in Computer Science},
  volume       = {2576},
  pages        = {268--289},
  publisher    = {Springer},
  year         = {2002},
  url          = {https://doi.org/10.1007/3-540-36413-7_20},
  doi          = {10.1007/3-540-36413-7_20},
  bibsource    = {dblp computer science bibliography, https://dblp.org}
}

@inproceedings{DBLP:conf/crypto/CamenischL04,
  author       = {Jan Camenisch and
                  Anna Lysyanskaya},
  editor       = {Matthew K. Franklin},
  title        = {Signature Schemes and Anonymous Credentials from Bilinear Maps},
  booktitle    = {Advances in Cryptology - {CRYPTO} 2004, 24th Annual International
                  CryptologyConference, Santa Barbara, California, USA, August 15-19,
                  2004, Proceedings},
  series       = {Lecture Notes in Computer Science},
  volume       = {3152},
  pages        = {56--72},
  publisher    = {Springer},
  year         = {2004},
  url          = {https://doi.org/10.1007/978-3-540-28628-8_4},
  doi          = {10.1007/978-3-540-28628-8_4},
  bibsource    = {dblp computer science bibliography, https://dblp.org}
}

@inproceedings{DBLP:conf/pkc/Connolly0P22,
  author       = {Aisling Connolly and
                  Pascal Lafourcade and
                  Octavio Perez{-}Kempner},
  editor       = {Goichiro Hanaoka and
                  Junji Shikata and
                  Yohei Watanabe},
  title        = {Improved Constructions of Anonymous Credentials from Structure-Preserving
                  Signatures on Equivalence Classes},
  booktitle    = {Public-Key Cryptography - {PKC} 2022 - 25th {IACR} International Conference
                  on Practice and Theory of Public-Key Cryptography, Virtual Event,
                  March 8-11, 2022, Proceedings, Part {I}},
  series       = {Lecture Notes in Computer Science},
  volume       = {13177},
  pages        = {409--438},
  publisher    = {Springer},
  year         = {2022},
  url          = {https://doi.org/10.1007/978-3-030-97121-2_15},
  doi          = {10.1007/978-3-030-97121-2_15},
  bibsource    = {dblp computer science bibliography, https://dblp.org}
}

@inproceedings{DBLP:conf/cans/BobolzEKRS21,
  author       = {Jan Bobolz and
                  Fabian Eidens and
                  Stephan Krenn and
                  Sebastian Ramacher and
                  Kai Samelin},
  editor       = {Mauro Conti and
                  Marc Stevens and
                  Stephan Krenn},
  title        = {Issuer-Hiding Attribute-Based Credentials},
  booktitle    = {Cryptology and Network Security - 20th International Conference, {CANS}
                  2021, Vienna, Austria, December 13-15, 2021, Proceedings},
  series       = {Lecture Notes in Computer Science},
  volume       = {13099},
  pages        = {158--178},
  publisher    = {Springer},
  year         = {2021},
  url          = {https://doi.org/10.1007/978-3-030-92548-2_9},
  doi          = {10.1007/978-3-030-92548-2_9},
  bibsource    = {dblp computer science bibliography, https://dblp.org}
}

@inproceedings{DBLP:conf/cms/LaponKDN11,
  author       = {Jorn Lapon and
                  Markulf Kohlweiss and
                  Bart De Decker and
                  Vincent Naessens},
  editor       = {Bart De Decker and
                  Jorn Lapon and
                  Vincent Naessens and
                  Andreas Uhl},
  title        = {Analysis of Revocation Strategies for Anonymous Idemix Credentials},
  booktitle    = {Communications and Multimedia Security, 12th {IFIP} {TC} 6 / {TC}
                  11 International Conference, {CMS} 2011, Ghent, Belgium, October 19-21,2011.
                  Proceedings},
  series       = {Lecture Notes in Computer Science},
  volume       = {7025},
  pages        = {3--17},
  publisher    = {Springer},
  year         = {2011},
  url          = {https://doi.org/10.1007/978-3-642-24712-5_1},
  doi          = {10.1007/978-3-642-24712-5_1},
  bibsource    = {dblp computer science bibliography, https://dblp.org}
}

@article{DBLP:journals/cacm/Chaum85,
  author       = {David Chaum},
  title        = {Security Without Identification: Transaction Systems to Make Big Brother
                  Obsolete},
  journal      = {Commun. {ACM}},
  volume       = {28},
  number       = {10},
  pages        = {1030--1044},
  year         = {1985},
  url          = {https://doi.org/10.1145/4372.4373},
  doi          = {10.1145/4372.4373},
  bibsource    = {dblp computer science bibliography, https://dblp.org}
}

@inproceedings{DBLP:conf/sp/Bonneau12,
  author       = {Joseph Bonneau},
  title        = {The Science of Guessing: Analyzing an Anonymized Corpus of 70 Million
                  Passwords},
  booktitle    = {{IEEE} Symposium on Security and Privacy, {SP} 2012, 21-23 May 2012,
                  San Francisco, California, {USA}},
  pages        = {538--552},
  publisher    = {{IEEE} Computer Society},
  year         = {2012},
  url          = {https://doi.org/10.1109/SP.2012.49},
  doi          = {10.1109/SP.2012.49},
  bibsource    = {dblp computer science bibliography, https://dblp.org}
}

@inproceedings{DBLP:conf/sp/SyversonGR97,
  author       = {Paul F. Syverson and
                  David M. Goldschlag and
                  Michael G. Reed},
  title        = {Anonymous Connections and Onion Routing},
  booktitle    = {1997 {IEEE} Symposium on Security and Privacy, May 4-7, 1997, Oakland,
                  CA, {USA}},
  pages        = {44--54},
  publisher    = {{IEEE} Computer Society},
  year         = {1997},
  url          = {https://doi.org/10.1109/SECPRI.1997.601314},
  doi          = {10.1109/SECPRI.1997.601314},
  bibsource    = {dblp computer science bibliography, https://dblp.org}
}

@inproceedings{DBLP:conf/diau/Raymond00,
  author       = {Jean{-}Fran{\c{c}}ois Raymond},
  editor       = {Hannes Federrath},
  title        = {Traffic Analysis: Protocols, Attacks, Design Issues, and Open Problems},
  booktitle    = {Designing Privacy Enhancing Technologies, International Workshop on
                  Design Issues in Anonymity and Unobservability, Berkeley, CA, USA,
                  July 25-26, 2000, Proceedings},
  series       = {Lecture Notes in Computer Science},
  volume       = {2009},
  pages        = {10--29},
  publisher    = {Springer},
  year         = {2000},
  url          = {https://doi.org/10.1007/3-540-44702-4_2},
  doi          = {10.1007/3-540-44702-4_2},
  bibsource    = {dblp computer science bibliography, https://dblp.org}
}

@inproceedings{DBLP:conf/crypto/Brands93,
  author       = {Stefan Brands},
  editor       = {Douglas R. Stinson},
  title        = {Untraceable Off-line Cash in Wallets with Observers (Extended Abstract)},
  booktitle    = {Advances in Cryptology - {CRYPTO} '93, 13th Annual International Cryptology
                  Conference, Santa Barbara, California, USA, August 22-26, 1993, Proceedings},
  series       = {Lecture Notes in Computer Science},
  volume       = {773},
  pages        = {302--318},
  publisher    = {Springer},
  year         = {1993},
  url          = {https://doi.org/10.1007/3-540-48329-2_26},
  doi          = {10.1007/3-540-48329-2_26},
  bibsource    = {dblp computer science bibliography, https://dblp.org}
}

@inproceedings{DBLP:conf/ccs/BrickellCC04,
  author       = {Ernest F. Brickell and
                  Jan Camenisch and
                  Liqun Chen},
  editor       = {Vijayalakshmi Atluri and
                  Birgit Pfitzmann and
                  Patrick D. McDaniel},
  title        = {Direct anonymous attestation},
  booktitle    = {Proceedings of the 11th {ACM} Conference on Computer and Communications
                  Security, {CCS} 2004, Washington, DC, USA, October 25-29, 2004},
  pages        = {132--145},
  publisher    = {{ACM}},
  year         = {2004},
  url          = {https://doi.org/10.1145/1030083.1030103},
  doi          = {10.1145/1030083.1030103},
  bibsource    = {dblp computer science bibliography, https://dblp.org}
}

@inproceedings{DBLP:conf/ccs/HanzlikS21,
  author       = {Lucjan Hanzlik and
                  Daniel Slamanig},
  editor       = {Yongdae Kim and
                  Jong Kim and
                  Giovanni Vigna and
                  Elaine Shi},
  title        = {With a Little Help from My Friends: Constructing Practical Anonymous
                  Credentials},
  booktitle    = {{CCS} '21: 2021 {ACM} {SIGSAC} Conference on Computer and Communications
                  Security, Virtual Event, Republic of Korea, November 15 - 19, 2021},
  pages        = {2004--2023},
  publisher    = {{ACM}},
  year         = {2021},
  url          = {https://doi.org/10.1145/3460120.3484582},
  doi          = {10.1145/3460120.3484582},
  bibsource    = {dblp computer science bibliography, https://dblp.org}
}

@techreport{I-D.kalos-bbs-blind-signatures,
  author = {Vasilis Kalos and Greg M. Bernstein},
  title = {Blind BBS Signatures},
  howpublished = {Working Draft},
  type = {Internet-Draft},
  number = {draft-kalos-bbs-blind-signatures-03},
  year = {2024},
  month = {October},
  institution = {IETF Secretariat},
}

@inproceedings{DBLP:conf/uss/HesseSS23,
  author       = {Julia Hesse and
                  Nitin Singh and
                  Alessandro Sorniotti},
  editor       = {Joseph A. Calandrino and
                  Carmela Troncoso},
  title        = {How to Bind Anonymous Credentials to Humans},
  booktitle    = {32nd {USENIX} Security Symposium, {USENIX} Security 2023, Anaheim,
                  CA, USA, August 9-11, 2023},
  pages        = {3047--3064},
  publisher    = {{USENIX} Association},
  year         = {2023},
  url          = {https://www.usenix.org/conference/usenixsecurity23/presentation/hesse},
  bibsource    = {dblp computer science bibliography, https://dblp.org}
}

@article{DBLP:journals/compsec/RodriguezKS24,
  author       = {Jes{\'{u}}s Garc{\'{\i}}a Rodr{\'{\i}}guez and
                  Stephan Krenn and
                  Daniel Slamanig},
  title        = {To pass or not to pass: Privacy-preserving physical access control},
  journal      = {Comput. Secur.},
  volume       = {136},
  pages        = {103566},
  year         = {2024},
  url          = {https://doi.org/10.1016/j.cose.2023.103566},
  doi          = {10.1016/J.COSE.2023.103566},
  bibsource    = {dblp computer science bibliography, https://dblp.org}
}

@article{DBLP:journals/popets/MirSBM23,
  author       = {Omid Mir and
                  Daniel Slamanig and
                  Balthazar Bauer and
                  Ren{\'{e}} Mayrhofer},
  title        = {Practical Delegatable Anonymous Credentials From Equivalence Class
                  Signatures},
  journal      = {Proc. Priv. Enhancing Technol.},
  volume       = {2023},
  number       = {3},
  pages        = {488--513},
  year         = {2023},
  url          = {https://doi.org/10.56553/popets-2023-0093},
  doi          = {10.56553/POPETS-2023-0093},
  bibsource    = {dblp computer science bibliography, https://dblp.org}
}

@misc{cryptoeprint:2024/2010,
      author = {Matteo Frigo and abhi shelat},
      title = {Anonymous credentials from {ECDSA}},
      howpublished = {Cryptology {ePrint} Archive, Paper 2024/2010},
      year = {2024},
      url = {https://eprint.iacr.org/2024/2010}
}

@misc{cryptoeprint:2024/2013,
      author = {Christian Paquin and Guru-Vamsi Policharla and Greg Zaverucha},
      title = {Crescent: Stronger Privacy for Existing Credentials},
      howpublished = {Cryptology {ePrint} Archive, Paper 2024/2013},
      url = {https://eprint.iacr.org/2024/2013},
      year = {2024}
}

@misc{Nitulescu2020zkSNARKsAG,
  title={zk-SNARKs: A Gentle Introduction},
  author={Anca Nitulescu},
  howpublished={\url{https://api.semanticscholar.org/CorpusID:211530704}},
  year={2020}
}

@misc{BeKaBBS23,
  title={BBS+ Applications, Standardization, and a Bit of Theory},
  author={Greg Bernstein and Vasilis Kalos},
  url={https://csrc.nist.gov/csrc/media/Presentations/2023/crclub-2023-10-18/images-media/20231018-crypto-club--greg-and-vasilis--slides--BBS.pdf},
  year={2023}

}

@article{SEC-030,
  url     = {http://dx.doi.org/10.1561/3300000030},
  year    = {2022},
  volume  = {4},
  journal = {Foundations and Trends® in Privacy and Security},
  title   = {Proofs, Arguments, and Zero-Knowledge},
  doi     = {10.1561/3300000030},
  issn    = {2474-1558},
  number  = {2–4},
  pages   = {117-660},
  author  = {Justin Thaler}
}

@inproceedings{cryptoeprint:2024/1846,
  author       = {Vadim Lyubashevsky and
                  Gregor Seiler and
                  Patrick Steuer},
  editor       = {Bo Luo and
                  Xiaojing Liao and
                  Jun Xu and
                  Engin Kirda and
                  David Lie},
  title        = {The LaZer Library: Lattice-Based Zero Knowledge and Succinct Proofs
                  for Quantum-Safe Privacy},
  booktitle    = {Proceedings of the 2024 on {ACM} {SIGSAC} Conference on Computer and
                  Communications Security, {CCS} 2024, Salt Lake City, UT, USA, October
                  14-18, 2024},
  pages        = {3125--3137},
  publisher    = {{ACM}},
  year         = {2024},
  url          = {https://doi.org/10.1145/3658644.3690330},
  doi          = {10.1145/3658644.3690330},
  bibsource    = {dblp computer science bibliography, https://dblp.org}
}

@inproceedings{DBLP:conf/stoc/BlumFM88,
  author       = {Manuel Blum and
                  Paul Feldman and
                  Silvio Micali},
  editor       = {Janos Simon},
  title        = {Non-Interactive Zero-Knowledge and Its Applications (Extended Abstract)},
  booktitle    = {Proceedings of the 20th Annual {ACM} Symposium on Theory of Computing,
                  May 2-4, 1988, Chicago, Illinois, {USA}},
  pages        = {103--112},
  publisher    = {{ACM}},
  year         = {1988},
  url          = {https://doi.org/10.1145/62212.62222},
  doi          = {10.1145/62212.62222},
  bibsource    = {dblp computer science bibliography, https://dblp.org}
}

@inproceedings{DBLP:conf/crypto/BootleLNS23,
  author       = {Jonathan Bootle and
                  Vadim Lyubashevsky and
                  Ngoc Khanh Nguyen and
                  Alessandro Sorniotti},
  editor       = {Helena Handschuh and
                  Anna Lysyanskaya},
  title        = {A Framework for Practical Anonymous Credentials from Lattices},
  booktitle    = {Advances in Cryptology - {CRYPTO} 2023 - 43rd Annual International
                  Cryptology Conference, {CRYPTO} 2023, Santa Barbara, CA, USA, August
                  20-24, 2023, Proceedings, Part {II}},
  series       = {Lecture Notes in Computer Science},
  volume       = {14082},
  pages        = {384--417},
  publisher    = {Springer},
  year         = {2023},
  url          = {https://doi.org/10.1007/978-3-031-38545-2_13},
  doi          = {10.1007/978-3-031-38545-2_13},
  bibsource    = {dblp computer science bibliography, https://dblp.org}
}

@inproceedings{DBLP:journals/iacr/ArgoGJLRS24,
  author       = {Sven Argo and
                  Tim G{\"{u}}neysu and
                  Corentin Jeudy and
                  Georg Land and
                  Adeline Roux{-}Langlois and
                  Olivier Sanders},
  editor       = {Bo Luo and
                  Xiaojing Liao and
                  Jun Xu and
                  Engin Kirda and
                  David Lie},
  title        = {Practical Post-Quantum Signatures for Privacy},
  booktitle    = {Proceedings of the 2024 on {ACM} {SIGSAC} Conference on Computer and
                  Communications Security, {CCS} 2024, Salt Lake City, UT, USA, October
                  14-18, 2024},
  pages        = {1523--1537},
  publisher    = {{ACM}},
  year         = {2024},
  url          = {https://doi.org/10.1145/3658644.3670297},
  doi          = {10.1145/3658644.3670297},
  bibsource    = {dblp computer science bibliography, https://dblp.org}
}

@inproceedings{DBLP:conf/scn/HebantP22,
  author       = {Chlo{\'{e}} H{\'{e}}bant and
                  David Pointcheval},
  editor       = {Clemente Galdi and
                  Stanislaw Jarecki},
  title        = {Traceable Constant-Size Multi-authority Credentials},
  booktitle    = {Security and Cryptography for Networks - 13th International Conference,
                  {SCN} 2022, Amalfi, Italy, September 12-14, 2022, Proceedings},
  series       = {Lecture Notes in Computer Science},
  volume       = {13409},
  pages        = {411--434},
  publisher    = {Springer},
  year         = {2022},
  url          = {https://doi.org/10.1007/978-3-031-14791-3_18},
  doi          = {10.1007/978-3-031-14791-3_18},
  bibsource    = {dblp computer science bibliography, https://dblp.org}
}

@inproceedings{DBLP:conf/ccs/MirBGLS23,
  author       = {Omid Mir and
                  Balthazar Bauer and
                  Scott Griffy and
                  Anna Lysyanskaya and
                  Daniel Slamanig},
  editor       = {Weizhi Meng and
                  Christian Damsgaard Jensen and
                  Cas Cremers and
                  Engin Kirda},
  title        = {Aggregate Signatures with Versatile Randomization and Issuer-Hiding
                  Multi-Authority Anonymous Credentials},
  booktitle    = {Proceedings of the 2023 {ACM} {SIGSAC} Conference on Computer and
                  Communications Security, {CCS} 2023, Copenhagen, Denmark, November
                  26-30, 2023},
  pages        = {30--44},
  publisher    = {{ACM}},
  year         = {2023},
  url          = {https://doi.org/10.1145/3576915.3623203},
  doi          = {10.1145/3576915.3623203},
  bibsource    = {dblp computer science bibliography, https://dblp.org}
}

@inproceedings{DBLP:conf/crypto/BelenkiyCCKLS09,
  author       = {Mira Belenkiy and
                  Jan Camenisch and
                  Melissa Chase and
                  Markulf Kohlweiss and
                  Anna Lysyanskaya and
                  Hovav Shacham},
  editor       = {Shai Halevi},
  title        = {Randomizable Proofs and Delegatable Anonymous Credentials},
  booktitle    = {Advances in Cryptology - {CRYPTO} 2009, 29th Annual International
                  Cryptology Conference, Santa Barbara, CA, USA, August 16-20, 2009.
                  Proceedings},
  series       = {Lecture Notes in Computer Science},
  volume       = {5677},
  pages        = {108--125},
  publisher    = {Springer},
  year         = {2009},
  url          = {https://doi.org/10.1007/978-3-642-03356-8_7},
  doi          = {10.1007/978-3-642-03356-8_7},
  bibsource    = {dblp computer science bibliography, https://dblp.org}
}

@article{DBLP:journals/istr/CamenischDELNPP14,
  author       = {Jan Camenisch and
                  Maria Dubovitskaya and
                  Robert R. Enderlein and
                  Anja Lehmann and
                  Gregory Neven and
                  Christian Paquin and
                  Franz{-}Stefan Preiss},
  title        = {Concepts and languages for privacy-preserving attribute-based authentication},
  journal      = {J. Inf. Secur. Appl.},
  volume       = {19},
  number       = {1},
  pages        = {25--44},
  year         = {2014},
  url          = {https://doi.org/10.1016/j.jisa.2014.03.004},
  doi          = {10.1016/J.JISA.2014.03.004},
  bibsource    = {dblp computer science bibliography, https://dblp.org}
}

@article{DBLP:journals/scn/Adams11,
  author       = {Carlisle Adams},
  title        = {Achieving non-transferability in credential systems using hidden biometrics},
  journal      = {Secur. Commun. Networks},
  volume       = {4},
  number       = {2},
  pages        = {195--206},
  year         = {2011},
  url          = {https://doi.org/10.1002/sec.136},
  doi          = {10.1002/SEC.136},
  bibsource    = {dblp computer science bibliography, https://dblp.org}
}

@techreport{I-D.kalos-bbs-per-verifier-linkability,
  author = {Vasilis Kalos and Greg M. Bernstein},
  title = {BBS per Verifier Linkability},
  howpublished = {Working Draft},
  type = {Internet-Draft},
  number = {draft-kalos-bbs-per-verifier-linkability-00},
  year = {2024},
  month = {October},
  institution = {IETF Secretariat},
}

@techreport{I-D.irtf-cfrg-bbs-signatures,
  author = {Tobias Looker and Vasilis Kalos and Andrew Whitehead and Mike Lodder},
  title = {The BBS Signature Scheme},
  howpublished = {Working Draft},
  type = {Internet-Draft},
  number = {draft-irtf-cfrg-bbs-signatures-07},
  year = {2024},
  month = {September},
  institution = {IETF Secretariat},
}

@article{DBLP:journals/popets/DavidsonGSTV18,
  author       = {Alex Davidson and
                  Ian Goldberg and
                  Nick Sullivan and
                  George Tankersley and
                  Filippo Valsorda},
  title        = {Privacy Pass: Bypassing Internet Challenges Anonymously},
  journal      = {Proc. Priv. Enhancing Technol.},
  volume       = {2018},
  number       = {3},
  pages        = {164--180},
  year         = {2018},
  url          = {https://doi.org/10.1515/popets-2018-0026},
  doi          = {10.1515/POPETS-2018-0026},
  bibsource    = {dblp computer science bibliography, https://dblp.org}
}

@inproceedings{DBLP:conf/uss/DingledineMS04,
  author       = {Roger Dingledine and
                  Nick Mathewson and
                  Paul F. Syverson},
  editor       = {Matt Blaze},
  title        = {Tor: The Second-Generation Onion Router},
  booktitle    = {Proceedings of the 13th {USENIX} Security Symposium, August 9-13,
                  2004, San Diego, CA, {USA}},
  pages        = {303--320},
  publisher    = {{USENIX}},
  year         = {2004},
  url          = {http://www.usenix.org/publications/library/proceedings/sec04/tech/dingledine.html},
  bibsource    = {dblp computer science bibliography, https://dblp.org}
}

@inproceedings{DBLP:conf/ccs/ChasePZ20,
  author       = {Melissa Chase and
                  Trevor Perrin and
                  Greg Zaverucha},
  editor       = {Jay Ligatti and
                  Xinming Ou and
                  Jonathan Katz and
                  Giovanni Vigna},
  title        = {The Signal Private Group System and Anonymous Credentials Supporting
                  Efficient Verifiable Encryption},
  booktitle    = {{CCS} '20: 2020 {ACM} {SIGSAC} Conference on Computer and Communications
                  Security, Virtual Event, USA, November 9-13, 2020},
  pages        = {1445--1459},
  publisher    = {{ACM}},
  year         = {2020},
  url          = {https://doi.org/10.1145/3372297.3417887},
  doi          = {10.1145/3372297.3417887},
  bibsource    = {dblp computer science bibliography, https://dblp.org}
}

@inproceedings{DBLP:conf/ccs/ChaseMZ14,
  author       = {Melissa Chase and
                  Sarah Meiklejohn and
                  Greg Zaverucha},
  editor       = {Gail{-}Joon Ahn and
                  Moti Yung and
                  Ninghui Li},
  title        = {Algebraic MACs and Keyed-Verification Anonymous Credentials},
  booktitle    = {Proceedings of the 2014 {ACM} {SIGSAC} Conference on Computer and
                  Communications Security, Scottsdale, AZ, USA, November 3-7, 2014},
  pages        = {1205--1216},
  publisher    = {{ACM}},
  year         = {2014},
  url          = {https://doi.org/10.1145/2660267.2660328},
  doi          = {10.1145/2660267.2660328},
  bibsource    = {dblp computer science bibliography, https://dblp.org}
}

@misc{cryptoeudi,
  author={Carsten Baum and
Olivier Blazy and
Jan Camenisch and
Jaap-Henk Hoepman  and
Eysa Lee and
Anja Lehmann and
Anna Lysyanskaya and
René Mayrhofer and
Hart Montgomery and
Ngoc Khanh Nguyen and
Bart Preneel and
shelat, a. and
Daniel Slamanig and
Stefano Tessaro and
Søren Eller Thomsen and
Carmela Troncoso},
  title = {{Cryptographers’ Feedback on the EU Digital Identity’s ARF}},
  howpublished = {\url{https://github.com/user-attachments/files/15904122/cryptographers-feedback.pdf}},
  year={2024}
}

@misc{anonterminology,
  author={Andreas Pfitzmann and Marit Hansen},
  title={Anonymity, Unlinkability, Unobservability, Pseudonymity, and
Identity Management – A Consolidated Proposal for Terminology}, 
  howpublished={\url{https://dud.inf.tu-dresden.de/literatur/Anon_Terminology_v0.28.pdf}},
  year={2006}
}
\else
 \printbibliography
\fi

\end{document}